\documentclass[fleqn,usenatbib]{mnras}

\usepackage{graphicx}
\usepackage{ifthen}
\usepackage{url}
\usepackage{amsmath}
\usepackage{bm}
\usepackage{lscape}
\usepackage{rotating}
\usepackage[graphicx]{realboxes}
\usepackage{supertabular}
\usepackage{pdfpages}
\usepackage{footnote}
\usepackage{hyperref}
\usepackage{amssymb}
\makesavenoteenv{tabular}
\makesavenoteenv{table}
\usepackage[T1]{fontenc}
\usepackage{tipa}
\usepackage{accents}
\newcommand{\ubar}[1]{\underaccent{\bar}{#1}}
\usepackage{tablefootnote}

\hypersetup{
  colorlinks   = true,    
  urlcolor     = blue,    
  linkcolor    = blue,    
  citecolor    = blue      
}

\usepackage{ulem}

\def\ltsima{$\; \buildrel < \over \sim \;$}
\def\lta{\lower.5ex\hbox{\ltsima}}
\def\gtsima{$\; \buildrel > \over \sim \;$}
\def\simgt{\lower.5ex\hbox{\gtsima}}
\def\kms{{\rm\,km \  s^{-1}}}

\def\kpc{{\rm\,kpc}}

\def\msun{{\rm\,M_\odot}}

\usepackage{amsmath}

\def\s{\ifmmode \widetilde \else \~\fi}
\def\={\overline}

\def\spose#1{\hbox to 0pt{#1\hss}}

\def\eg{{e.g.,\,}}
\def\ie{{i.e.,\,}}
\def\lta{\mathrel{\spose{\lower 3pt\hbox{$\mathchar"218$}}
     \raise 2.0pt\hbox{$\mathchar"13C$}}}
\def\gta{\mathrel{\spose{\lower 3pt\hbox{$\mathchar"218$}}
     \raise 2.0pt\hbox{$\mathchar"13E$}}}
\def\Dt{\spose{\raise 1.5ex\hbox{\hskip3pt$\mathchar"201$}}}    
\def\dt{\spose{\raise 1.0ex\hbox{\hskip2pt$\mathchar"201$}}}    

\def\dotsfill{\leaders\hbox to 1em{\hss.\hss}\hfill}

\def\FeH{{\rm[Fe/H]}}
\def\MgFe{{\rm[Mg/Fe]}}

\def\ione{{~\sc i}}
\def\ii{{~\sc ii}}

\loadboldmathitalic 
\title[Extreme outskirts of Ursa Minor]{The extended "stellar halo" of the Ursa Minor dwarf galaxy} 
\author[F. Sestito et al.]{Federico Sestito$^{1}$\thanks{Email: \url{sestitof@uvic.ca}}, Daria Zaremba$^{1,2}$, 
Kim A. Venn$^{1}$,
Lina D'Aoust$^{1}$,
Christian Hayes$^{3}$,
\newauthor
Jaclyn Jensen$^{1}$,
Julio F. Navarro$^{1}$,
Pascale Jablonka$^{4,5}$,
Emma Fern\'andez-Alvar$^{6,7}$,
\newauthor
Jennifer Glover$^{1}$,
Alan W. McConnachie$^{3,1}$ ,
and Andr\'e-Nicolas Chen\'e$^{8,9}$
\\
$^{1}$ Department of Physics and Astronomy, University of Victoria, PO Box 3055, STN CSC, Victoria BC V8W 3P6, Canada\\
$^{2}$ National University of Kyiv-Mohyla Academy, 04655 Kyiv, Ukraine\\
$^{3}$ NRC Herzberg Astronomy \& Astrophysics, 5071 West Saanich Road, Victoria, BC V9E 2E7, Canada\\
$^{4}$ Laboratoire d'astrophysique, \'Ecole Polytechnique F\'ed\'erale de Lausanne (EPFL), Observatoire, CH-1290 Versoix, Switzerland\\
$^{5}$ GEPI, Observatoire de Paris, Universit\'e PSL, CNRS, 5 Place Jules Janssen, F-92195 Meudon, France\\
$^{6}$ Instituto de Astrofısica de Canarias, E-38200 La Laguna, Tenerife, Spain\\
$^{7}$ Dept. Astrofısica, Universidad de La Laguna, E-38206 La Laguna, Tenerife, Spain\\
$^{8}$ Gemini Observatory/NSF’s NOIRLab, 670 N. A'ohoku Place, Hilo, Hawai'i, 96720, USA\\
$^{9}$ Visiting astronomer at the Universit\'e de Montr\'eal, Complexe des Sciences, Montr\'eal, QC H2V 0B3, Canada
}
\pubyear{2023}
\date{Accepted XXX. Received YYY; in original form ZZZ}

\begin{document}
\maketitle 
\label{firstpage}
\pagerange{\pageref{firstpage}--\pageref{lastpage}}

\begin{abstract}
Stellar candidates in the Ursa Minor (UMi) dwarf galaxy have been found using a new Bayesian algorithm applied to  \textit{Gaia} EDR3 data. 
Five of these targets are located in the extreme outskirts of UMi, from $\sim5$ to 12 elliptical half-light radii (r$_h$), where r$_h$(UMi) $= 17.32 \pm 0.11$ arcmin, and 
have been observed with the GRACES high resolution spectrograph at the Gemini-Northern telescope. Precise  radial velocities ($\sigma_{\rm{RV}} < 2$ km s$^{-1}$) and metallicities ($\sigma_{\rm{\FeH}} < 0.2$ dex) confirm their memberships of UMi. Detailed analysis of the brightest and outermost star (Target~1, at $\sim12$ r$_h$), yields precision chemical abundances for the $\alpha$- (Mg, Ca, Ti), odd-Z (Na, K, Sc), Fe-peak (Fe, Ni, Cr), and neutron-capture (Ba) elements. With data from the literature and APOGEE DR17, we find the chemical patterns in UMi are consistent with an outside-in star formation history that includes yields from core collapse supernovae, asymptotic giant branch stars, and supernovae Ia. Evidence for a knee in the [$\alpha$/Fe] ratios near $\FeH\sim-2.1$ indicates a low star formation efficiency similar to that in other dwarf galaxies. Detailed analysis of the surface number density profile shows evidence that UMi's outskirts have been populated by tidal effects, likely as a result of completing multiple orbits around the Galaxy.
\end{abstract}

\begin{keywords}
stars: abundances -- stars: Population II --
galaxies : formation -- galaxies: dwarf -- galaxies: individual: Ursa Minor -- galaxies: evolution
\end{keywords}

\section{Introduction}
$\Lambda-$Cold Dark Matter ($\Lambda-$CDM)  predicts that massive galaxies grow from the accretion of smaller systems \citep[\eg][]{White78,Frenk88,NavarroFrenkWhite97}.  Therefore, galaxies are expected to be surrounded by extended "stellar halos" built from disrupted systems \citep[\eg][]{Helmi08}. A stellar halo is clearly observed in large galaxies like the Milky Way, but they remain elusive and poorly studied in dwarf galaxies \citep[][and references therein]{Deason22}. One reason is likely that the fraction of stellar mass assembled through mergers is reduced at the dwarf galaxy mass scales, where "in-situ" star formation dominates \citep[\eg][]{Genel10}. 

Given their shallow gravitational potential, faint dwarf galaxies are also susceptible to internal processes, such as star formation and the subsequent stellar feedback \citep[\eg][]{ElBadry18b}; and external, such as mergers \citep[\eg][]{Deason14}, ram pressure stripping \citep[\eg][]{Grebel03} and stirring \citep[\eg][]{Kazantzidis11}, tidal interaction \citep[\eg][]{Fattahi18}, and reionization \citep[\eg][]{Wheeler19}. All of these processes may act to influence their individual morphologies \citep[\eg][and references therein]{Higgs21}.
Signatures of these mechanisms will be most evident in the outskirts ($\gtrsim 4$ r$_h$) of a dwarf galaxy, where accreted remnants can show up as an excess of stars over and above expectations from a simple single-component model (akin to a stellar halo in a more massive galaxy).

Only in the past few years, with the advent of the exquisite Gaia astrometric and photometric data, it has become possible to find stars in the extreme outskirts of dwarf galaxies. \citet{Chiti21} identified member stars up to $\sim$9 half-light radii ($r_h$) away from the centre of the faint dwarf galaxy, Tucana II, suggesting that the outskirts originated from a merger or a bursty stellar feedback; \citet{Filion21} and \citet{Longeard22} analysed the chemo-dynamical properties of Bo{\"o}tes~I, proposing that the system might have been more massive in the past and that tidal stripping has affected the satellite; \citet{Yang22} analysed the extent of the red giant branch of Fornax and identified a break in the density distribution, which they interpret as the presence of an extended stellar halo up to a distance of $7$ half-light radii; \citet{Longeard23} found new members in Hercules up to $\sim10$ half-light radii, and noted that the lack of a strong velocity gradient argued against ongoing tidal disruption; \citet{Sestito23scl} found new Sculptor members up to 10 r$_h$, and proposed that the system is perturbed by tidal effects; \citet{Waller23} discussed that the chemistry of the outermost stars in Coma Berenices, Ursa Major~I, and Bo{\"o}tes~I  is consistent with their formation in the central regions, then moving them to their current locations, maybe through tidal stripping and/or supernovae feedback.

In this paper, we explore the outer most regions of Ursa Minor (UMi).
UMi is historically a well-studied system. The system is at the low-end of the classical dwarf galaxies in terms of stellar mass \citep[$\sim2.9\cdot10^5$ M$_{\odot}$, \eg][]{Mcconnachie12,Simon19}. Some controversies remain  regarding the star formation history (SFH) and its efficiency. For example, \citet{Carrera02} suggested that up to  $\sim$95 per cent of UMi stars are older than 10 Gyr, invoking an episodic SFH at early times. This is based on studies of its colour-magnitude diagram \citep[\eg][]{Mighell99,Bellazzini02}. Other models interpreted the chemical properties of UMi as due to extended SFH, from 3.9 and 6.5 Gyr \citep{Ikuta02,Ural15}. \citet{Kirby11,Kirby13} matched the wide metallicity distribution function (MDF) of UMi with a chemical evolution model that includes infall of gas. 

In addition, \citet{Ural15} developed three chemical evolution models, showing that winds from supernovae are needed to describe UMi's MDF, especially to reproduce stars at  higher metallicities. The authors underline that winds help to explain the absence of gas at the present time. In agreement with \citet{Ikuta02}, their models use an extended low-efficiency SFH duration \citep[5 Gyr,][]{Ural15}. 

Finally, the $\Lambda-$CDM  cosmological zoom-in simulations  developed by \citet{Revaz18} found that the star formation and chemical evolution of UMi can be explained. In particular, when SNe~Ia and II events are taken into account with thermal blastwave-like feedback \citep[][and references therein]{Revaz18}, then they can reproduce the observed distribution in  metallicity,  \MgFe{}, and the radial velocity dispersion invoking a short star formation of only 2.4 Gyr.  

From high-resolution GRACES spectroscopy of 5 outermost stars in UMi, we revisit the
chemo-dynamical evolution of this classical dwarf galaxy.
Our results,  combined with spectroscopic results for additional stars in the literature, are used to discuss the extended chemical and dynamical evolution of UMi. The target selection, the observations, and the spectral reduction are reported in Section~\ref{sec:data}. Stellar parameters are inferred in Section~\ref{sec:stellparam}. The model atmosphere and chemical abundance analysis for the most distant UMi star (Target~1) are reported in Section~\ref{sec:modelatmo}~and~\ref{Sec:chemabu}, respectively. Section~\ref{Sec:ewcat} describes the measurement of \FeH{} using Ca Triplet lines for Target~2--5.  The chemo-dynamical properties of Ursa Minor are discussed in Section~\ref{sec:discussion}.
Appendix~\ref{sec:orbpar} reports the inference of the orbital parameters of UMi.

\section{Data}\label{sec:data}

\subsection{Target selection}\label{sec:selection}
A first selection of candidate member stars for spectroscopic follow-up is made using the algorithm described in \citet{Jensen23}. Similarly to its predecessor, described in \citet{McVenn2020a}, this algorithm is designed to search for member stars in a given dwarf galaxy by determining a probability of membership to the satellite. The probability of being a satellite member, P$_{\rm{sat}}$, is composed by three likelihoods based on the system’s (1) colour-magnitude diagram (CMD), (2) systemic proper motion, and (3) the projected radial distance from the center of the satellite, using precise data from \textit{Gaia} EDR3 \citep[][]{GaiaEDR3}. A notable improvement to the algorithm in \citet{Jensen23} is the model for the spatial likelihood. The stellar density profile of a dwarf can often be approximated by a single exponential function \citep[see][]{McVenn2020a}. In order to search for tidal features or extended stellar haloes, the spatial likelihood in \citet{Jensen23} assumes that each system may host a secondary, extended, and lower density, outer profile. Only a handful of systems are found in their work to host an outer profile, a few of which are already known to be perturbed by tidal effects (\eg Bootes III and Tucana III). Also shown in their work, Ursa Minor is a system for which a secondary outer profile is observed, indicating either an extended stellar halo or tidal features.
This algorithm has proved useful to identify new members in the extreme outskirts of some ultra-faint and classical dwarf galaxies \citep{Waller23,Sestito23scl} and effectively   removes Milky Way foreground contamination. 

We selected stars with a high probability ($>80$\%) of being associated to UMi, and at a distance greater than 5 half-light radii ($\gtrsim 85$ arcmin or $\gtrsim 2$ kpc) from the centre of the dwarf. This included five red giants with magnitudes in the range $17.4 \leq G \leq 18.3$ mag in the \textit{Gaia} EDR3 G band. The brightest target is also the farthest in projection, reaching an extreme distance of 11.7 half-light radii from the centre of UMi.  Our other four targets, at a distance of $5.2-6.3 \ \ r_h$, are also listed as highly likely UMi candidates by \citet[][with a probability $>90$ percent]{Qi22}. The main properties of UMi and our five targets are reported in Tables~\ref{tab:umiprop} and \ref{tab:targets}, respectively.

\begin{figure*}
\includegraphics[width=1\textwidth]{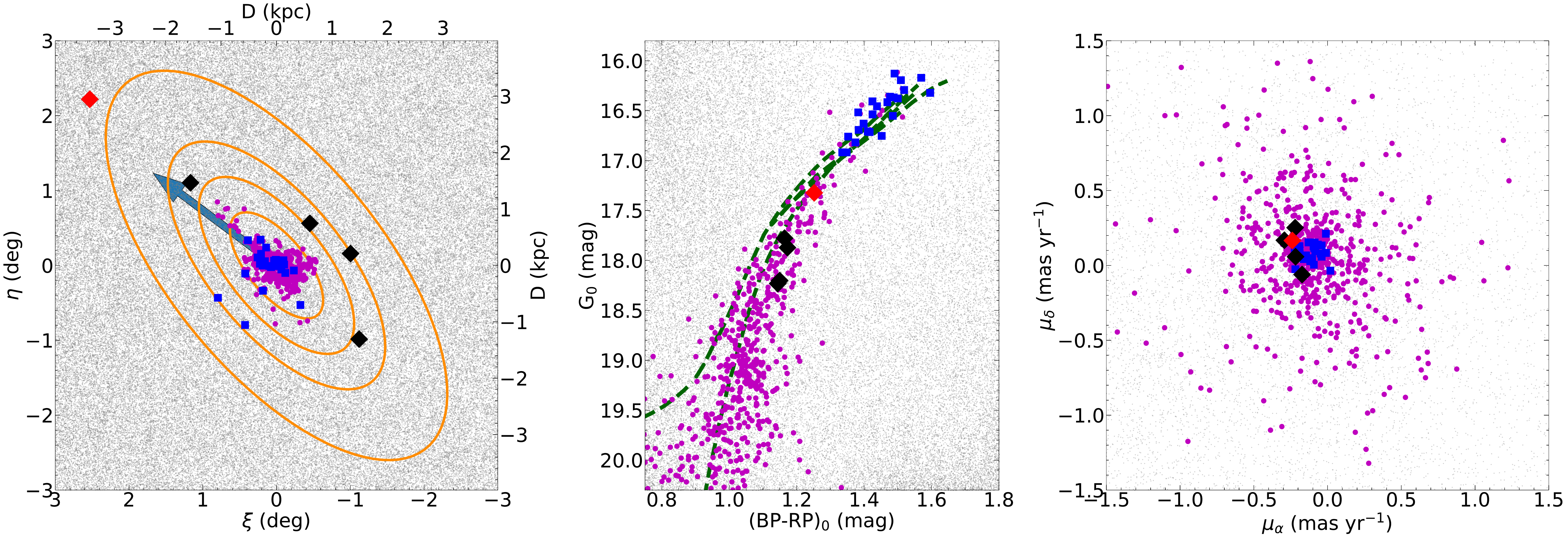}
\caption{Ursa Minor seen through \textit{Gaia} EDR3. All the panels: Target~1 is marked with a red diamond, while black diamonds are Target~2--5. Magenta circles are UMi literature stars from \citet{Spencer18} and \citet{Pace20}. Blue squares are UMi stars selected from APOGEE DR17. MW foreground stars are marked with grey small dots. These are selected from \textit{Gaia} EDR3 in the direction of UMi and within the field of view of the $\eta-\xi$ panel. Left panel: Projected sky coordinates and projected distance from UMi centre. The orange ellipses denotes the elliptical distances from UMi centre of 3, 5, 7, and 11 $r_h$. The arrow points in the direction of UMi proper motion. Central panel: Colour-magnitude diagram. Dark green dashed lines is a Padova isochrone at $\FeH = -2.0$ and age of 12 Gyr \citep{Bressan12}. Right panel: Proper motion space.}
\label{Fig:onsky}
\end{figure*}

The position of our five candidates, together with other known UMi members, are shown in Figure~\ref{Fig:onsky} in  projected sky coordinates, on the colour-magnitude diagram, and in  proper motion space. This shows  
that our algorithm is able to select new candidate members even in the very outskirts of the system.  

We also gather UMi members from \citet{Spencer18}, \citet{Pace20}, and from the APOGEE data release 17 \citep[DR17,][]{APOGEEDR17} for Figure~\ref{Fig:onsky}, and cross-match with Gaia EDR3 to retrieve coordinates, proper motion, and photometry. Our selection algorithm was also applied to the APOGEE DR17 targets to select stars with high membership probability ($>70$ \%) and high signal-to-noise in their spectra (SNR $>70$). Surprisingly, two stars from APOGEE DR17 have an elliptical distance of $\sim7$ $r_h$. The \FeH{} values for those two stars are at the edge of the metallicity grid of APOGEE ($\FeH \sim -2.4$); thus, while their radial velocity measurements are precise, their true \FeH{} could be lower, in turn  affecting their [X/Fe] ratios.

\begin{table}
\caption[]{Galactic parameters of Ursa Minor. The coordinates $\alpha,\delta$, the mean metallicity, the  mean radial velocity, the velocity dispersion, the heliocentric distance  D$_\odot$, the ellipticity, the position angle $\phi$, and the  half-light radius  r$_{\rm h}$ in arcmin and pc, the mean proper motion from Gaia EDR3, the dynamical mass, the mass density, and the luminosity are reported with the respective references. (a) refers to \citet{Mcconnachie12}, (b) to \citet{McVenn2020a}, (c) to \citet{McVenn2020b}, (d) to \citet{Qi22}, and (e) to \citet{Mateo98}.}
\centering
\resizebox{0.47\textwidth}{!}{
\hspace{-0.6cm}
\begin{tabular}{lrc}
\hline
Property & Value & Reference\\
\hline
$\alpha$ & $227.2854$ deg  & (b) \\
 $\delta$ & $67.2225$ deg & (b) \\
 $\overline{\FeH}$ & $-2.13\pm0.01$ & (b)\\
  $\overline{\rm{RV}}$ & $246.9\pm 0.1$ km s$^{-1}$ & (b)\\
   $\sigma_{\rm{V}}$ & $9.5\pm 1.2$ km s$^{-1}$ & (b)\\
 D$_\odot$   & $76\pm10$ kpc & (a) \\
ellipticity  &  $0.55\pm0.01$ & (b) \\
 $\phi$   &  $50\pm1 $ deg & (b) \\
  r$_{\rm h}$ & $17.32\pm0.11$ arcmin  & (b) \\
  r$_{\rm h}$ & $382\pm 53$ pc & (b) \\ 
r$_{\rm h,plummer}$ & 407 pc & (d) \\ 
  $\mu_{\alpha}\rm{cos}\delta$ & $-0.124 \pm 0.004$ mas yr$^{-1}$  & (c)\\
   $\mu_{\delta}$ & $0.078 \pm 0.004$ mas yr$^{-1}$  & (c) \\
    M$_{\rm{dyn}}(\leq \rm{r}_{\rm{half}})$ & $9.5\times10^6\msun$& (a)\\
    Mass density & $0.35 \msun$ pc$^{-3}$ & (e) \\
    L & $0.29\times 10^6$ L$_{\odot}$ & (e)  \\
\hline \hline
\end{tabular}}
\label{tab:umiprop}
\end{table}

\begin{table*}
\caption[]{The \textit{Gaia} EDR3 source ID, the coordinates $(\alpha,\delta)$, the projected coordinates $(\xi,\eta)$, the elliptical radius distance $r_{\rm{ell}}$ in $r_h$ unit, the probability to be a member from \citet{Jensen23}, the \textit{Gaia} EDR3 photometry G and BP$-$RP, and the reddening A$_{\rm V}$ from \citet{Schlafly11} are reported for each target.}
\centering
\resizebox{1\textwidth}{!}{
\hspace{-0.6cm}
\begin{tabular}{lcccccccccc}
\hline
Target &  source id & $\alpha$ & $\delta$ & $\xi$  & $\eta$  & $r_{\rm{ell}}$ & $P_{\rm{sat}}$ & G & BP$-$RP & A$_{\rm V}$ \\
 & & (deg)& (deg) & (deg) & (deg) & ($r_h$) & & (mag) & (mag)& (mag)   \\ \hline
 Target~1  & $1647329728514964352$ & $234.45303$ & $69.29204$ & $2.53226$ & $2.21888$ & $11.67$ &  $0.80$ & $17.39$ &  $1.29$ &  $0.08$  \\ 
Target~2  & $1693464785444020224$ & $224.67731$ & $67.35983$ & $-1.00378$ & $0.15842$ & $6.34$ &  $0.97$ & $17.83$ &  $1.19$ &  $0.06$  \\ 
Target~3  & $1693573430936780032$ & $226.08983$ & $67.77965$ & $-0.45214$ & $0.56153$ & $5.55$ &  $0.96$ & $17.91$ &  $1.19$ &  $0.05$  \\ 
Target~4  & $1669324938936435200$ & $224.50756$ & $66.21361$ & $-1.12033$ & $-0.98413$ & $5.17$ &  $0.94$ & $18.25$ &  $1.17$ &  $0.06$  \\ 
Target~5  & $1645948119139534336$ & $230.43949$ & $68.29581$ & $1.16629$ & $1.10328$ & $5.60$ &  $0.92$ & $18.29$ &  $1.17$ &  $0.06$  \\ 

\hline
\end{tabular}}
\label{tab:targets}
\end{table*}

\subsection{GRACES observations}
The five targets were observed with the Gemini Remote Access to CFHT ESPaDOnS Spectrograph \citep[GRACES,][]{Chene14,Pazder14} using the 2-fibre (object+sky) mode with a resolution of R$\sim40000$. GRACES consists a 270-m optical fibre that connects the Gemini North telescope to the Canada–France–Hawaii Telescope ESPaDOnS cross-dispersed high resolution \'echelle spectrograph \citep{Donati06}.  The spectral coverage of GRACES is from 4500 \AA{} to 10000 \AA{} \citep{Chene14}. The targets were observed as part of  the GN-2022A-Q-128 program  (P.I. F. Sestito).

For the brightest target (Target~1, G$=17.4$ mag), which is also the farthest one from the centre ($\sim 11.7 \ \ r_h$), we obtained a spectrum with SNR per resolution element of $\sim30$ at the Ba\ii{} 6141 \AA{} region. This spectrum has sufficient SNR to measure the abundances for additional elements, specifically the $\alpha-$ (Mg, Ca, Ti), odd$-$Z (Na, K, Sc), Fe$-$peak (Fe, Cr, Ni), and neutron$-$capture process (Ba) elements across the entire GRACES spectral coverage. We refer to this observational set-up as the ``high-SNR mode''. For the remaining four targets, which have distances from 5 $-$ 7 $r_h$, a SNR per resolution element of $\sim20$ in the Ca\ii{} T region ($\sim$8550 \AA) was obtained for precise radial velocities and metallicities. In this ``low-SNR mode'', the metallicities are derived from the equivalent width (EW) of the NIR Ca\ii{} T, as described in Section~\ref{Sec:ewcat}. Observing information is summarized in Table~\ref{tab:obs}, including the signal-to-noise ratio measured at the Mg\ione{} b, Ba\ii{} 614nm, and Ca\ii{} T regions.

\begin{table}
\caption[]{Total exposure time, number of exposures,  signal-to-noise ratio (SNR) measured at the Mg\ione{} 518nm, Ba\ii{} 614nm, and Ca\ii{} 850nm regions, and the observation dates are reported for each target. The SNR is defined as the ratio between the median flux and its standard deviation in given spectral region.
}

\resizebox{0.48\textwidth}{!}{
\hspace{-0.6cm}
\begin{tabular}{lcccccc}
\hline
Target  & $t_{\rm exp}$ & N$_{\rm exp}$ & SNR & SNR & SNR & Obs. date\\ 
  & (s) &  & @Mg\ione b & @Ba\ii & @Ca\ii T & YY/MM/DD \\ \hline
Target~1  & 14400 & 6 &9 & 27& 37 & 22/06/18 \\
Target~2  &1800 &1 & 5 & 12  & 17& 22/03/14 \\
Target~3  & 1800& 1& 1& 6  & 8& 22/03/14 \\
Target~4  &2400 &1 & 2 & 6  & 11 & 22/06/17 \\
Target~5  &2400 & 1& 1& 5 &10& 22/06/17 \\
\hline
\end{tabular}}
\label{tab:obs}
\end{table}

\subsection{Spectral reductions}
The GRACES spectra were first reduced using the Open source Pipeline for ESPaDOnS Reduction and Analysis \citep[OPERA,][]{Martioli12} tool, which also corrects for heliocentric motion. Then the reduced spectra were post-processed following an updated  procedure of the pipeline described in \citet{Kielty21}. The latter pipeline allows us to measure the radial velocity of the observed star, to co-add  multiple observations, to check for possible radial velocity variations, to correct for the motion of the star, and to eventually re-normalise the flux. This procedure also improves the signal-to-noise ratio in the overlapping spectral order regions without downgrading the spectral resolution. Radial velocities are reported in Table~\ref{tab:params}.

This procedure failed for one of the spectral orders of Target~1 covering the Mg\ione{} b region for reasons that we could not overcome within the scope of this project. We therefore extracted the data for Target~1 ourselves using \textsc{DRAGraces}\footnote{\url{https://github.com/AndreNicolasChene/DRAGRACES/releases/tag/v1.4}} \textsc{IDL} code \citep{Chene21}.

\begin{figure*}
\includegraphics[width=1\textwidth]{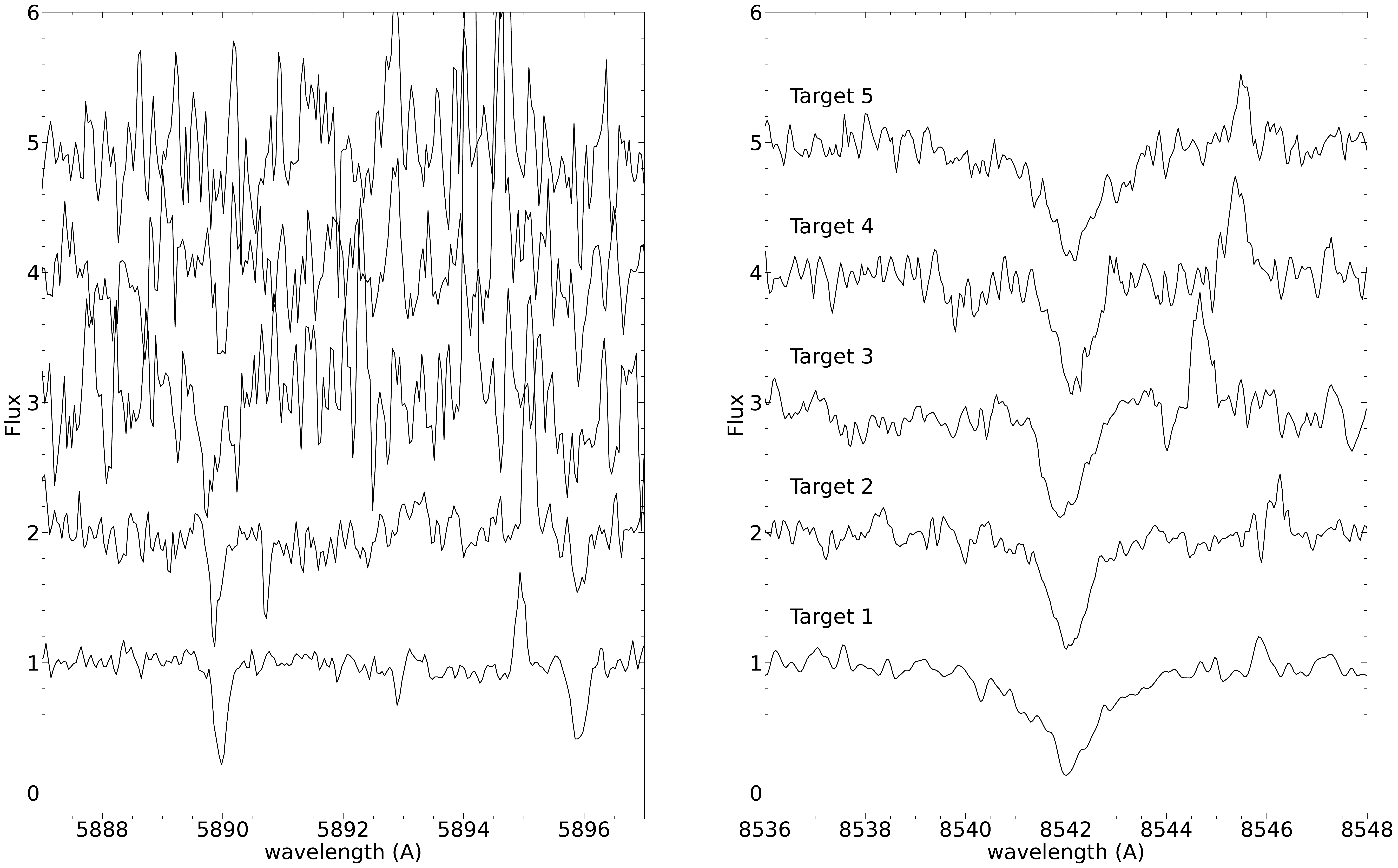}
\caption{GRACES spectra for the five new UMi member stars. Left panel: Na\ione{} Doublet region. Chemical abundance ratios are measurable only for Target~1 given the low SNR of Targets~2--5. Right panel: The second component of the Ca\ii{} Triplet. This spectral line is used to infer [Fe/H] (see Section~\ref{Sec:ewcat}).}
\label{Fig:spectra}
\end{figure*}

The final spectra for all five targets near the Na\ione{} Doublet (left panel) and in the NIR Ca\ii{} Triplet (right panel) regions are shown in Figure~\ref{Fig:spectra}. The quality of the spectra indicates that the adopted exposure time were sufficient for the requested science, \ie chemical abundances for Target~1, and \FeH{} and RV only for Targets~2$-$5.

\section{Stellar parameters}\label{sec:stellparam}
Given the low SNR of our spectra, we use the InfraRed flux method (IRFM) from \citet{Gonzalez09} with photometry from  {\it Gaia} EDR3 to find the effective temperatures, adopting the \citet[][]{Mucciarelli21} colour-temperature relationship for giants. The input parameters are the {\it Gaia} EDR3 (BP $-$ RP) de-reddened colour and a metallicity estimate. The 2D \citet{Schlafly11} map\footnote{\url{https://irsa.ipac.caltech.edu/applications/DUST/}} has been used to correct the photometry for extinction\footnote{To convert from the  E(B-V) map to  {\it Gaia} extinction coefficients,  the  $\rm A_V/E(B-V)= 3.1$ \citep{Schultz75} and the $\rm A_G/A_V = 0.85926$, $\rm A_{BP} /A_V = 1.06794$, $\rm A_{RP} /A_V = 0.65199$ relations \citep{Marigo08,Evans18} are used.}. As input metallicities, we adopt the value $\FeH=-2.0\pm0.5$, compatible with the metallicity distribution in UMi.

Surface gravities were found using the Stefan-Boltzmann equation\footnote{$L_{\star} = 4\pi R_{\star}^2 \sigma T_{\star}^4$; the radius of the star can be calculated from this equation, then the surface gravity is inferred assuming the mass.}. This step required the effective temperature, the distance of the object, the {\it Gaia} EDR3 G de-reddened photometry, and the bolometric corrections on the flux \citep{Andrae18} as input.  A Monte Carlo algorithm has been applied to the input parameters with their uncertainties to estimate the total uncertainties on the stellar parameters. The input quantities were randomised within $1\sigma$ using a  Gaussian distribution, except for the stellar mass.  The latter is treated with a flat prior from 0.5 to 0.8 $\msun$, which is consistent with the mass of  long-lived very metal-poor stars. The mean uncertainty on the effective temperature is $\sim 94$~K, while on the surface gravity it is $\sim 0.08$~dex. This method has been shown to provide reliable stellar parameters suitable for spectroscopic studies of very metal-poor stars \citep[\eg][]{Kielty21,Sestito23,Waller23}. The  stellar parameters are reported in Table~\ref{tab:params}.

\begin{table}
\caption[]{Stellar parameters of the five targets. \FeH{} for Target~1 is from Fe\ione{} and Fe\ii{} lines, while for the other stars is from Ca\ii{} Triplet lines.}
\resizebox{0.49\textwidth}{!}{
\hspace{-0.6cm}
\begin{tabular}{lcccc}
\hline
Target &  RV & T$_{\rm{eff}}$ & log~g & \FeH  \\ 
 & (km s$^{-1})$ & (K)&  &    \\ \hline 
 Target~1  &  $-256.91 \pm 0.05$ & $4604 \pm 94$ & $1.15 \pm 0.08$ & $-2.09 \pm 0.09$ \\ 
Target~2  &  $-265.26 \pm 1.89$ & $4771 \pm 93$ & $1.43 \pm 0.07$ & $-2.80 \pm 0.34$ \\ 
Target~3  &  $-218.78 \pm 1.82$ & $4760 \pm 100$ & $1.45 \pm 0.08$ & $-2.67 \pm 0.31$ \\ 
Target~4  &  $-245.63 \pm 1.78$ & $4795 \pm 85$ & $1.60 \pm 0.07$ & $-2.85 \pm 0.32$ \\ 
Target~5  &  $-247.29 \pm 1.63$ & $4814 \pm 100$ & $1.61 \pm 0.08$ & $-2.31 \pm 0.37$ \\ 
\hline
\end{tabular}}
\label{tab:params}
\end{table}

\section{Model atmospheres analysis - Target 1}\label{sec:modelatmo}
In this Section, we describe the model atmospheres, the method, and the atomic data for our spectral line list adopted to determine detailed chemical abundances for Target~1.

\subsection{Model atmospheres}
Model atmospheres are generated from the \textsc{MARCS}\footnote{\url{https://marcs.astro.uu.se}} models \citep{Gustafsson08,Plez12}; in particular, we selected the  \textsc{OSMARCS} spherical models as Target~1 is a giant with log(g)$<3.5$. An initial set of model atmospheres was generated by varying the derived stellar parameters and metallicity $\FeH=-2.0$ within their uncertainties, and adopting a microturbulence velocity ($v_t =2.02 \kms$) scaled by the surface gravity from the calibration by \citet{Mashonkina17} for giants.

\subsection{The lines list and the atomic data}
Spectral lines were selected from our previous analyses of very metal-poor stars in the Galactic halo and other nearby dwarf galaxies observed with GRACES \citep{Norris17,Monty20,Kielty21}. Atomic data is taken from \textsc{linemake}\footnote{\url{https://github.com/vmplacco/linemake}} \citep{Placco21}, with the exception of K\ione{} lines taken from the National Institute of Standards and Technology \citep[NIST,][]{NIST_ASD}\footnote{NIST database at \url{https://physics.nist.gov/asd}}.

\subsection{Spectral line measurements}
Spectral line measurements are made using spectrum synthesis, broadened with a Gaussian smoothing kernel of FWHM = 0.15, which matches the resolution of the GRACES 2-fibre mode spectra) in a four-step process: (1) the synthesis of the \FeH{} lines in our  initial line list (see above) is carried out using an initial model atmosphere and invoking the \textsc{MOOG}\footnote{\url{https://www.as.utexas.edu/~chris/moog.html}} spectrum synthesis program \citep{Sneden73,Sobeck11};  (2) a new \FeH{} is determined by removing noisy lines; (3) the set of model atmospheres is updated with the new \FeH{} as metallicity; (4) the chemical abundances are derived using the updated model atmospheres and our full line list. The final chemical abundance is given by the average measurement in case of  multiple spectral lines.  
 
\subsection{Checking the stellar parameters}
Excitation equilibrium in the line abundances of Fe\ione{} is a check on the quality of the effective temperature. For Target~1, the slope in A(Fe\ione) $-$ Excitation potential (EP) from the linear fit has a value of $ -0.027$ dex eV$^{-1}$. This value is smaller than the dispersion in the measurements of the chemical abundances ($\sim0.2$ dex) over the range in EP ($\sim$4 eV). Thus, we conclude our effective temperature estimates are sufficient from the IRFM.

Ionization balance between Fe\ione{} $-$ Fe\ii{} is widely used as a sanity check on the surface gravity estimates \citep[\eg][]{Mashonkina17}. However, \citet{Karovicova20} have strongly advised against using this method for very metal-poor giants. They used interferometric observations of metal-poor stars to find radii, and subsequently precise  stellar parameters for a set of metal-poor benchmark stars. With their stellar parameters, they have found that deviations in Fe\ione{} $-$ Fe\ii{} can reach up to $\sim0.8$ dex. This effect is the strongest in very metal-poor cool giants (\eg \FeH$<-2.0$, log(g)$<3$, and T$_{\rm eff}\lesssim 5500$ K), such as UMi Target~1 (see Table~\ref{tab:params}).   If we examine A(Fe\ione{}) and A(Fe\ii{}) in UMi Target~1, we find they differ by only  $1.43 \sigma$ or $0.16\pm0.11$ dex.  This value is consistent with ionization equilibrium, and also  within the range in the discrepancies found by  \citet{Karovicova20} for cool giants. For these reasons, we refrain from tuning the surface gravity based on the Fe lines.

\section{Chemical abundance analysis - Target 1 }\label{Sec:chemabu}
This section describes the chemical abundances that we determine from the spectrum of Target~1.
This includes an application of non-local thermodynamic equilibrium corrections, and a comparison with other UMi members and MW halo stars in the literature.

\subsection{$\alpha-$elements}
\begin{figure}
\includegraphics[width=0.5\textwidth]{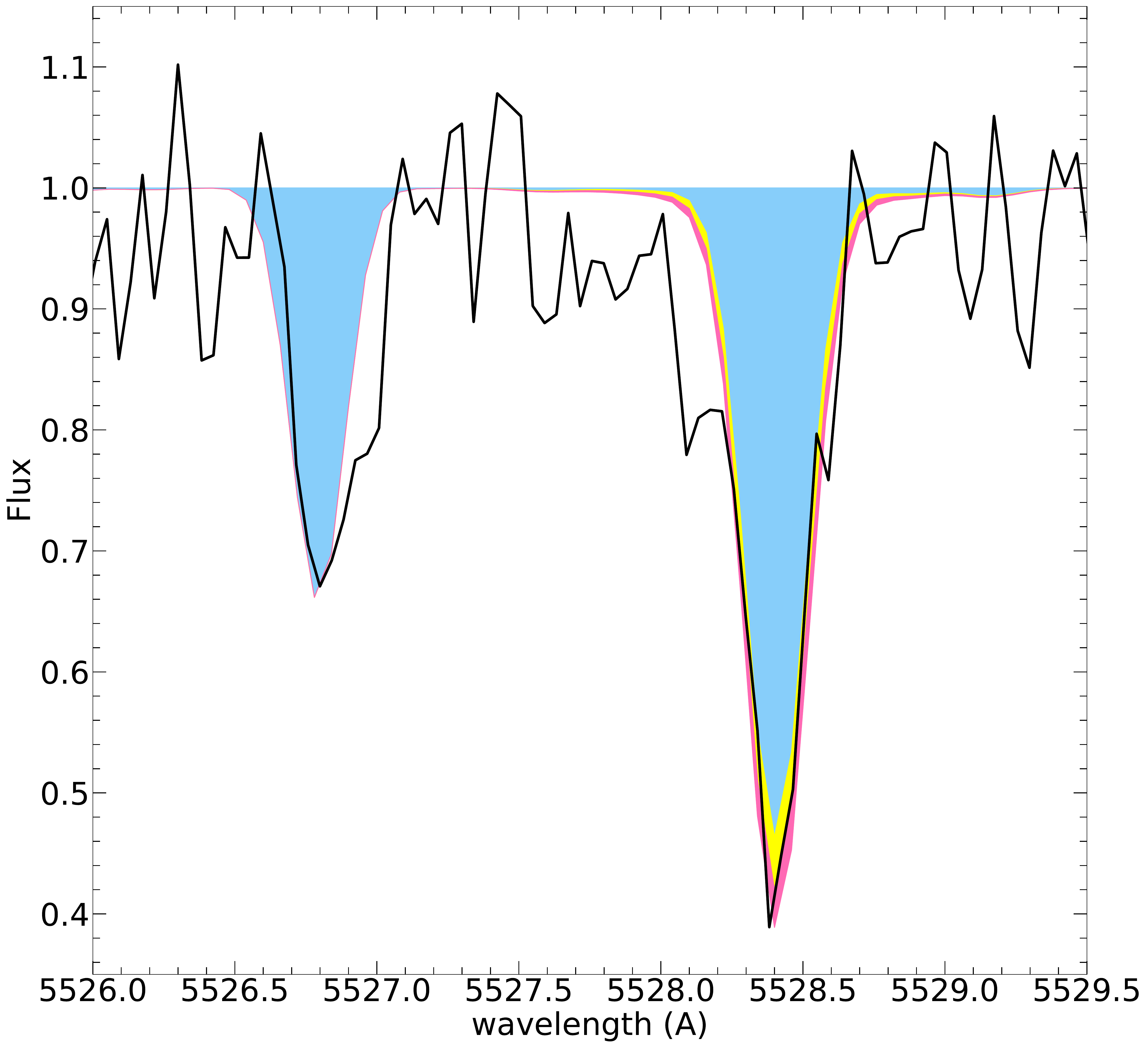}
\caption{Mg\ione{} 5528\AA{} region. The Mg-rich spectrum of Target~1 (black solid line) is compared with three synthetic spectra with \MgFe{} $=+0.5, +0.8, +1.0$ (light blue, yellow, and pink shaded areas, respectively). Synthetic spectra have been generated using the \textit{synth} mode in \textsc{MOOG} \citep{Sneden73} with the line list from \textsc{linemake} \citep{Placco21}. Model atmosphere are from  \textsc{MARCS}  \citep{Gustafsson08,Plez12}. The synthetics are created at the same resolution of GRACES and with the stellar parameters and metallicity as Target~1.}
\label{Fig:mgcheck}
\end{figure}

\begin{figure*}
\includegraphics[width=1.\textwidth]{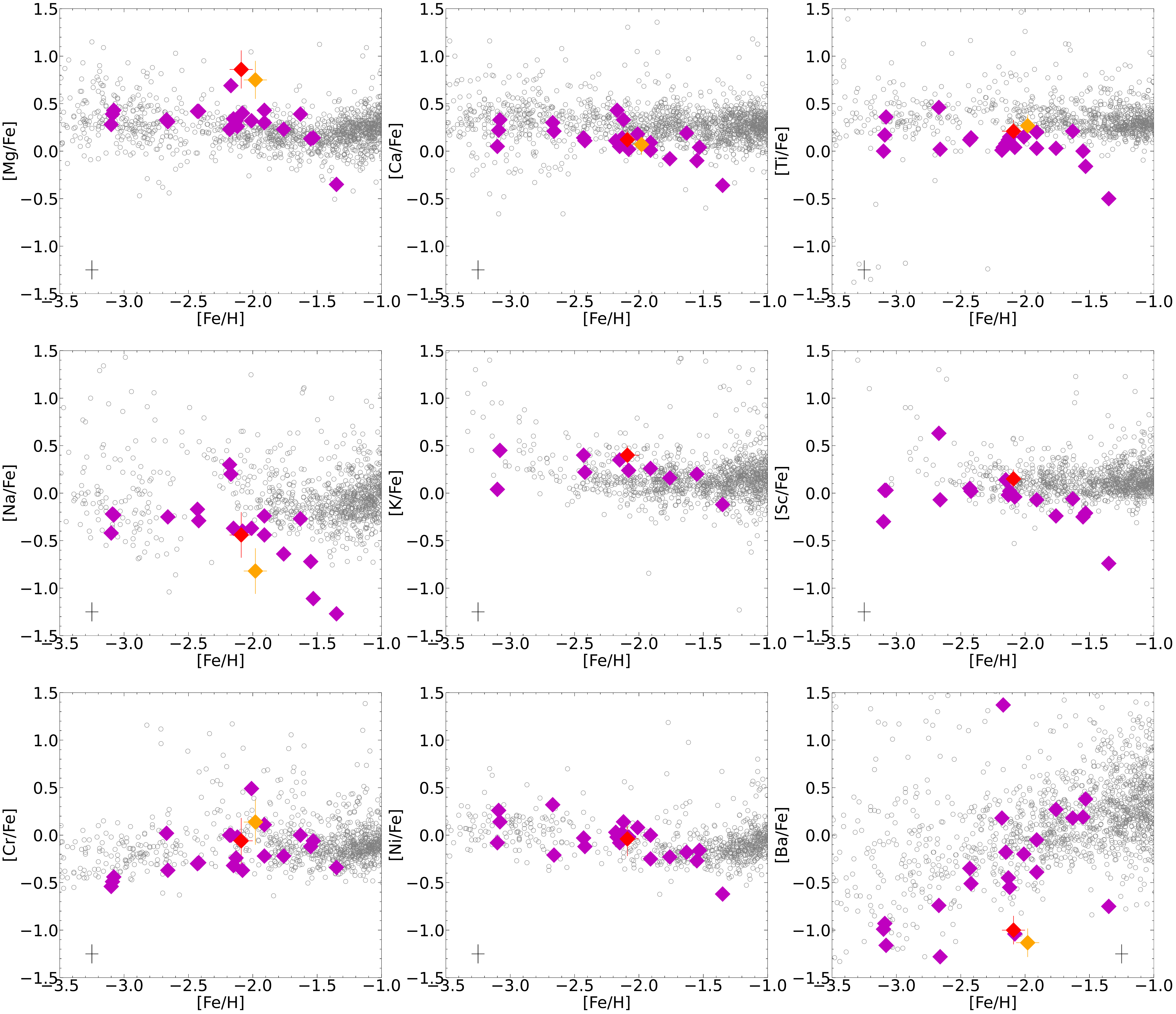}
\caption{Chemical abundances for stars in UMi. Target~1 is marked with a red diamond (LTE) and with an orange diamond (NLTE). UMi stars from the high-resolution observations from literature are denoted with magenta diamonds. The literature compilation is from \citet{Shetrone01}, \citet{Sadakane04}, \citet{Cohen10}, \citet{Kirby12}, and \citet{Ural15} and it is in LTE. Grey open circles mark MW halo stars compiled from \citet{Aoki13}, \citet{Yong13}, \citet{Kielty21}, and \citet{Buder21}. The black cross at the corner of each panel represents the typical uncertainty on the UMi literature chemical abundances.}
\label{Fig:chems}
\end{figure*}

$\alpha$-elements are primarily formed in the cores of massive stars and during the explosive phases of core-collapse supernovae \citep[\eg][]{Timmes95,Kobayashi20}. There are only three $\alpha$-elements with measurable lines in our GRACES spectrum of Target~1; Mg, Ca, Ti. The A(Mg\ione{}) is from two lines of the Mg\ione{} Triplet ($\lambda\lambda 5172.684, 5183.604$\AA) and the weaker $5528.405$\AA{} line.  We display the strong Mg lines in Target~1 against three synthetic spectra with [Mg/Fe] $= +0.5, +0.8, +1.0$ in Figure~\ref{Fig:mgcheck}. The A(Ca\ione{}) is inferred from 13 spectral lines, from 5588 \AA{} to 6500 \AA.  Up to 12 and 9 lines of Ti\ione{} and Ti\ii{} are useful to infer A(Ti\ione) and A(Ti\ii), respectively. The first row of panels in Figure~\ref{Fig:chems} display the [Mg, Ca, Ti/Fe] ratios as a function of the \FeH{}. Both the LTE and NLTE analysis are reported (see Section~\ref{nltesec}). Since both Ti\ione{} and Ti\ii{} lines are present in the spectrum, [Ti/Fe] is the average weighted by the number of lines of each species.

\subsection{Odd-Z elements}
Odd-Z elements are excellent tracers of metal-poor core-collapse supernovae due to the odd-even effect in the predicted yields \citep{Heger10, Nomoto13, Kobayashi20, Ebinger20}. Three odd-Z elements are observable in our spectrum of Target~1; Na, K, Sc. Sodium is measurable from the spectral lines of the Na\ione{} Doublet ($\lambda\lambda 5889.951,5895.924$ \AA). K\ione{} is observable with two lines at $\lambda\lambda7664.899, 7698.965$ \AA{}. These lines are very close to water vapour lines of the Earth's atmosphere; however, the radial velocity for Target~1 places these lines in clear windows. Sc is measured from only one Sc\ii{} line at $\lambda\lambda5526.785$ \AA{}. The abundances of K and Sc have been measured with the  \textsc{synth} configuration in \textsc{MOOG}, taking into account hyperfine splitting effects for Sc. The second row of panels of Figure~\ref{Fig:chems} shows  [Na, K, Sc/Fe] (LTE for all and also NLTE for Na).

\subsection{Fe-peak elements}
Fe-peak elements are important tracers of stellar evolution. At early times, they were produced primarily in core collapse supernovae and then later in supernovae Ia events \citep[e.g.,][]{Heger10, Nomoto13}. The Fe-peak elements observable in our GRACES spectra include Fe, Cr and Ni.  The abundance from Fe I is from 29 lines, while A(Fe\ii) is from only 3 lines. Our final [Fe/H] values are the average measurements weighted by the number of lines per star. Chromium is measured from 3 spectral lines of Cr I ($\lambda\lambda$ 5296.691, 5345.796, 5409.783 \AA), while Nickel is found from four lines Ni I lines ($\lambda\lambda$  5476.904, 5754.656, 6586.31, 6643.63 \AA). The left and centre panels of the third row of Figure~\ref{Fig:chems} show [Cr/Fe] (LTE and NLTE) and [Ni/Fe] (LTE) as a function of \FeH{}.

\subsection{Neutron-capture process elements}
Neutron-capture elements are primarily synthesised through two main channels, the rapid and the slow neutron captures processes. If the neutron capture timescale is shorter than the $\beta^{-}$ decay time, then rapid-process elements are formed. Conditions where this is most likely to happen are found in core collapse supernovae and neutron-star mergers. Otherwise, as in the stellar atmospheres of AGB stars, where neutron fluxes are lower and have weaker energies, then the beta-decay timescale is shorter, leading to the production via the slow-neutron capture processes. The only neutron-capture process element present in our GRACES spectra is Ba, with two Ba\ii{} lines ($\lambda\lambda 6141.73, 6496.91 $ \AA. To infer the A(Ba\ii{}), \textsc{MOOG} has been run with the synthetic configuration to account for the hyperfine structure corrections. Bottom right panel of Figure~\ref{Fig:chems} displays [Ba/Fe] (LTE and NLTE) as a function of \FeH{}.

\subsection{NLTE corrections}\label{nltesec}
The elemental abundances in the atmospheres of very metal-poor stars are affected by  departures from Local Thermodynamic Equilibrium (LTE). Thus, the statistical equilibrium solutions need to be corrected for radiative effects (non-LTE effects, or ``NLTE''), which can be large for some species. To correct for  NLTE effects in Fe \citep{Bergemann2012} and Na\ione{} \citep{Lind2012}, we adopted the results compiled in the \textsc{INSPECT}\footnote{\url{http://inspect-stars.com}} database. The NLTE corrections for Mg\ione{} \citep{Bergemann2017}, Ca\ione{} \citep{Mashonkina17}, Ti\ione{} and Ti\ii{} \citep{Bergemann2011}, and Cr\ione{} \citep{Bergemann2010b} are from the MPIA webtool database\footnote{\url{http://nlte.mpia.de}}. For Ba\ii{} lines, we adopted the NLTE corrections from \citet{Mashonkina19}, also available online\footnote{\url{http://www.inasan.ru/\~lima/pristine/ba2/}}.

\begin{table}
\caption[]{Chemical abundances of Target~1. The LTE and NLTE ratios are reported together with the $\sigma$ and the number of lines for each measured species. For Fe and ti we report the number of lines relative to both the neutral and the single-ionised states.}
\resizebox{0.48\textwidth}{!}{
\hspace{-0.6cm}
\begin{tabular}{lrrrr}
\hline
Ratio &  LTE & $\sigma$ & N$_{\rm{lines}}$ & NLTE  \\ 
 & (dex) & (dex)& & (dex)   \\ \hline 
[Fe/H] &  $-2.09$ & $0.09$ & 29$+$3 & $-1.98$  \\ 
 \rm{[Mg/Fe]} &  0.86 & 0.20 & 3 & 0.75  \\ 
 \rm{[Ca/Fe]} &  0.12 & 0.11 & 13 & 0.07  \\ 
  \rm{[Ti/Fe]} &  0.21 &  0.12 & 12$+$9 & 0.27  \\ 
  \rm{[Na/Fe]} &  $-0.44$ & $0.24$ & 2 & $-0.82$  \\ 
 \rm{[K/Fe] }&  0.40 & 0.10 &2 & $--$  \\ 
 \rm{[Sc/Fe]}&0.15 & 0.10 &1 &  $--$ \\ 
 \rm{[Cr/Fe]} &  $-0.06$ & 0.24 & 3 & 0.14  \\ 
 \rm{[Ni/Fe]} &  $-0.04$ & 0.18 & 4 & $--$  \\ 
 \rm{[Ba/Fe] }&  $-1.00$ & 0.15 &2 & $-1.13$  \\ 
\hline
\end{tabular}}
\label{tab:chems}
\end{table}

\subsection{Uncertainty on the chemical abundances}
The  uncertainty  on element X is given by $\sigma_{\rm A(X)}=\delta_{\rm A(X)}/\sqrt{{\rm N_X}}$ if the number of the measured spectral lines is ${\rm N_X}>5$, or  $\sigma_{\rm A(X)}=\delta_{\rm A(Fe\ione)}/\sqrt{{\rm N_X}}$ otherwise.  The terms $\delta_{\rm A(X)}$ and $\delta_{\rm A(Fe\ione)}$ include the errors due to uncertainties in the stellar parameters (see Table~\ref{tab:params}). Given the SNR across the observed combined spectrum of Target~1, the uncertainty on the chemical abundance ratios is in the range $0.10 \leq \sigma_{\rm{[X/Fe]}}\leq 0.24$. This range for the uncertainty is compatible with the ones measured by \citet{Kielty21} and \citet{Waller23}, in which they use a similar observational setup with GRACES to study  chemical abundances of very metal-poor giant stars.

\subsection{Elemental abundance compilation from the literature}\label{sec:chemslit}
UMi is an interesting and nearby dwarf galaxy that has had extensive observations of stars in its inner regions. We have gathered the elemental abundance results from optical high-resolution observations of stars in  Ursa Minor from the literature, shown for comparisons in Figure~\ref{Fig:chems}. This compilation is composed of 21 stars in total, including \citet[][4 stars]{Shetrone01}, \citet[][3 stars]{Sadakane04}, \citet[][10 stars]{Cohen10}, \citet[][1 star]{Kirby12}, and \citet[][3 stars]{Ural15}. All of these studies provide 1D LTE chemical abundances.

We also compare the chemistry of the stars in UMi with those in the MW halo from a compilation including \citet{Aoki13,Yong13,Kielty21,Buder21}. The stars from \citet{Buder21} are from the third data release of GALactic Archaeology with HERMES \citep[GALAH,][]{DeSilva15,Buder21} collaboration. We select GALAH stars to be in the halo, with reliable metallicities (\textsc{flag\_fe = 0}), chemical abundances (\textsc{flag\_X\_fe = 0}), and stellar parameters (\textsc{flag\_sp = 0}).

\section{Metallicities from the NIR Ca\ii{} T lines}\label{Sec:ewcat}

For Targets~2--5 observed in low-SNR mode, metallicities are derived from the NIR Ca\ii{} T lines. We follow the method described in \citet{Starkenburg10} with some minor modifications. Starting with their Equation~A.1: 
\begin{equation}
    \FeH = a+b\cdot \rm{M_V} + c\cdot \rm{EW_{2+3}} + d\cdot \rm{EW_{2+3}^{-1.5}} + e\cdot \rm{EW_{2+3}} \cdot \rm{M_V},
    \label{Eq:ewmet}
\end{equation}
where $\rm{M_V}$ is the absolute V magnitude of the star, $\rm{EW_{2+3}}$ is the sum of the equivalent width of the Ca\ii{} $\lambda\lambda 8542.09,8662.14$ \AA{} lines, and $a,b,c,d$ are the coefficients listed in Table~A.1 of \citet{Starkenburg10}. $\rm{M_V}$ is derived converting the \textit{Gaia} EDR3 magnitudes to the Johnson-Cousin filter following the relation from \citet[][see their Table C.2 for the coefficients]{Riello21} and adopting a heliocentric distance of $76\pm10$ \kpc{} \citep[\eg][]{Mcconnachie12}. Our minor modification is due to the fact that the third component of our Ca\ii{} T spectra is contaminated by sky lines. Therefore, $\rm{EW_{2+3}}$ is derived assuming that the EW ratio between the second and the third Ca\ii{} T lines is $\rm{EW_{8542}}/\rm{EW_{8662}}=1.21\pm0.03$, in agreement with \citet[][see their Figure~B.1]{Starkenburg10}. The EW of the Ca\ii{} 8542 \AA{} line is measured using the \textsc{splot} routine in \textsc{IRAF} \citep{Tody86,Tody93}, fitting the line with multiple profiles. The median and the standard deviation have been adopted as final values for the EW and its uncertainty. We perform a Monte Carlo test with $10^6$ randomisations on the heliocentric distance, the $\rm{EW_{8542}}$, the $\rm{EW_{8542}}/\rm{EW_{8662}}$ ratio, and the de-reddened magnitudes assuming a Gaussian distribution.  The final \FeH{} and its uncertainty are the median and the standard deviation from the randomisations, respectively.

Although \citet{Starkenburg10} proved that this metallicity calibration is reliable and compatible with high-resolution studies, we use Target~1 to check for a possible offset in \FeH. Given the different SNR between Target~1 ($\sim35$ at Ca\ii{} T) and the other targets ($\sim8-15$ at Ca\ii{} T), the spectrum of Target~1 has been degraded to match the SNR of the other targets. Its metallicity from  Ca\ii{} T is $\FeH_{\rm{CaT}}=-2.22\pm0.36$, compatible within $0.35\sigma$ with the metallicity inferred from Fe lines ($\FeH=-2.09\pm0.09$). The SNR of the Ca\ii{} T region in the observed spectra is sufficient to obtain an uncertainty on the metallicity of $\sim0.33$ dex. Table~\ref{tab:params} reports the inferred metallicities together with the stellar parameters and radial velocities.

\section{Discussion}\label{sec:discussion}

In this section, we discuss the membership of the five targets observed with GRACES, the chemical evolution, and the chemo-dynamical properties of the dwarf galaxy Ursa Minor.

\subsection{Five new distant members of UMi}\label{sec:newmembers}

\begin{figure*}
\includegraphics[width=\textwidth]{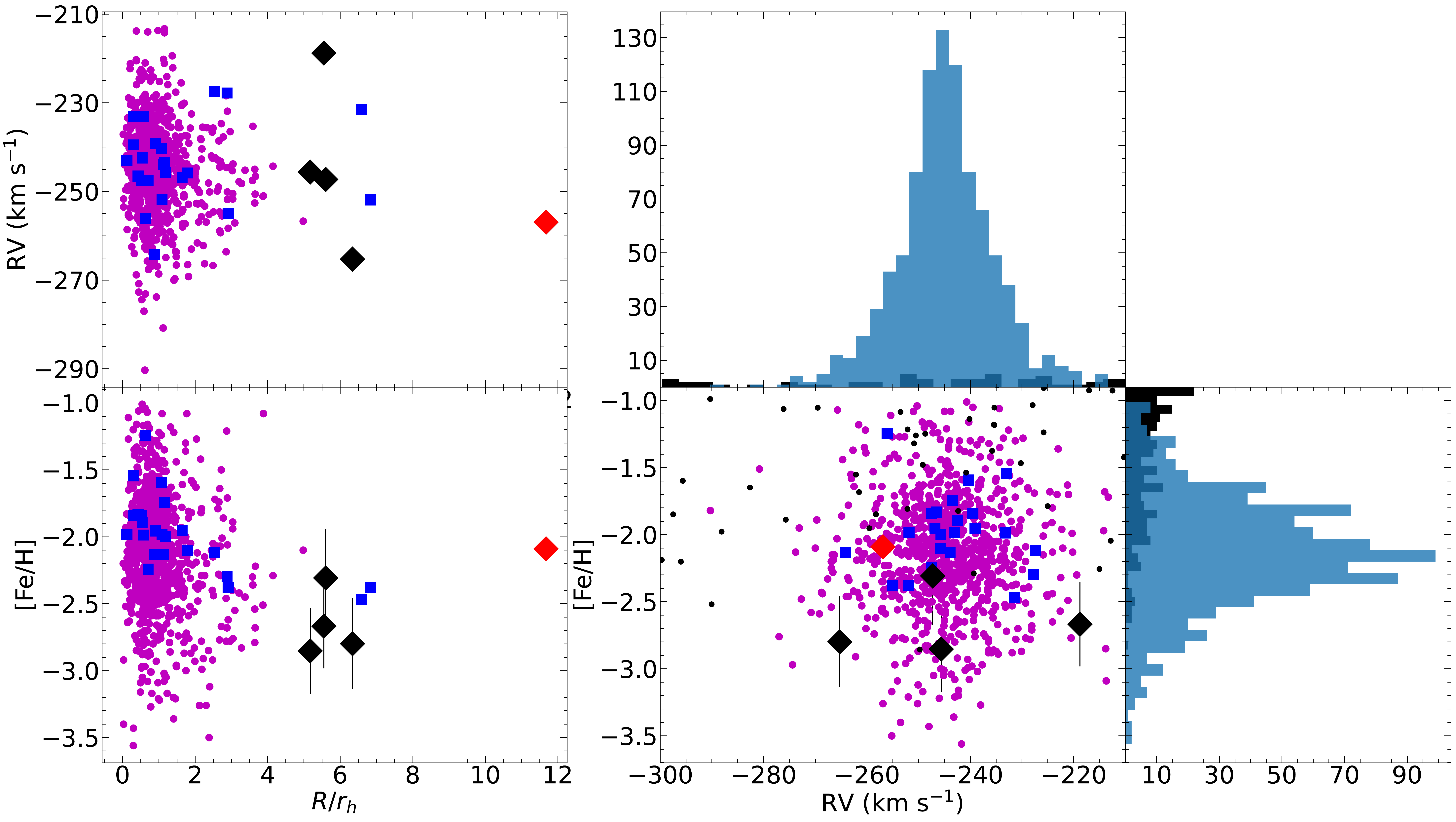}
\caption{Distribution of UMi stars. Left panels: Radial velocities (top) and metallicities (bottom) as a function of the elliptical distance. Central panel: distribution of UMi stars in the \FeH{} vs. RV space and of the Besan\c{c}on stellar particles (black dots). Corner plots: histograms of the RV (top) and metallicities (right) distributions of UMi star in blue. Besan\c{c}on simulations are displayed in black. Target~1 is marked with a red diamond, while Target~2--5 are displayed with black diamonds. Magenta dots are the compilation of stars from  \citet{Spencer18} and \citet{Pace20}. Blue squares are  UMi members selected from APOGEE DR17.}
\label{Fig:rv_met_rh}
\end{figure*}

Radial velocities and metallicities were measured for five new targets in UMi from GRACES spectra, where  \FeH{} is from Fe\ione{} and Fe\ii{} lines in case of Target~1 (see Section~\ref{Sec:chemabu}), while \FeH{} is inferred through the NIR Ca\ii{} Triplet lines  for Target~2--5 (see Section~\ref{Sec:ewcat}). Figure~\ref{Fig:rv_met_rh} displays the metallicities and radial velocities of our targets and known UMi members \citep[][APOGEE DR17]{Spencer18,Pace20} as a function of their elliptical distances (left panels); the \FeH{} vs. RV space and their histograms (central and right panels). The five targets  have  metallicities and  radial velocities compatible with the UMi distributions, therefore we identify them as new members of UMi. At a first glance, stars in the outer region ($r_{\rm{ell}} \gtrsim 4 r_h$) seem to have a larger RV dispersion, but, it is consistent within $0.8\sigma$ to the  RV dispersion in the central regions.

To further exclude the possibility that our UMi targets are halo interlopers, we examine the Besan\c{c}on simulation of the MW halo \citep{Robin03,Robin17}. Star particles are selected in the UMi direction (\ie same field-of-view as the left panel of Figure~\ref{Fig:onsky}) and nearby stars are removed (heliocentric distance $\leq5$ kpc). This leaves to 300 star particles. Only 39 particles inhabit the same RV $-$ \FeH{} range in the right panels of Figure~\ref{Fig:rv_met_rh} (displayed by black dots). Within this smaller sample, three star particles have the same proper motion as UMi; however, the photometry of these three star particles is brighter by 2 magnitudes in the G band from members of UMi at the same colour BP $-$ RP. Therefore, the Besan\c{c}on MW halo simulation fails to reproduce our observed quantities for stars in UMi (spatial coordinates, proper motion, CMD, RV, and \FeH{}). This supports our conclusion that Targets~1--5 are new UMi members, and not foreground stars. We notice from Figure~\ref{Fig:chems} that Target 1 stands out in [Ba/Fe] with unusually low Ba for a star with similar \FeH{} in the MW. This is a very rare occurrence in the MW, which strengthen the hypothesis that Target~1 is a new UMi member.

\subsection{Chemical Evolution of UMi}

\subsubsection{The Chemistry of Target~1}\label{sec:ccsne}

The detailed chemistry of Target~1 may provide a glimpse into the early star formation events in UMi, depending on its age and how it was moved to the outermost regions of this satellite galaxy. 
One possibility is that it may have formed just after the contributions from SNe~II and was exiled by early supernovae feedback and/or tidal forces from pericentric passage(s) with the Galaxy (see Section~\ref{sec:tides}).

Target 1 has a low [Ba/Fe] and it is also lower in [Na, Ca/Mg] (anticipating Figure~\ref{Fig:pisne}) than the other stars in UMi and the MW (Figure~\ref{Fig:chems}). A similar abundance pattern has been found in some stars in Coma Berenices \citep{Frebel12}, Segue 1 \citep{Frebel14}, Hercules \citep{Koch08,Koch13,Francois16}, and in the Milky Way \citep[\eg][]{Sitnova19,Kielty21,Sestito23}. This has been interpreted as contribution from only one or a few early core-collapse SNe~II (CCSNe), known as the ``one-shot'' model \citep{Frebel12}.  To test this hypothesis, we explore a variety of low mass, low metallicity CCSN models to compare their predicted yields to our chemical abundances in Target~1.

\begin{figure}
\hspace{-0.3cm}\includegraphics[width=0.51\textwidth]{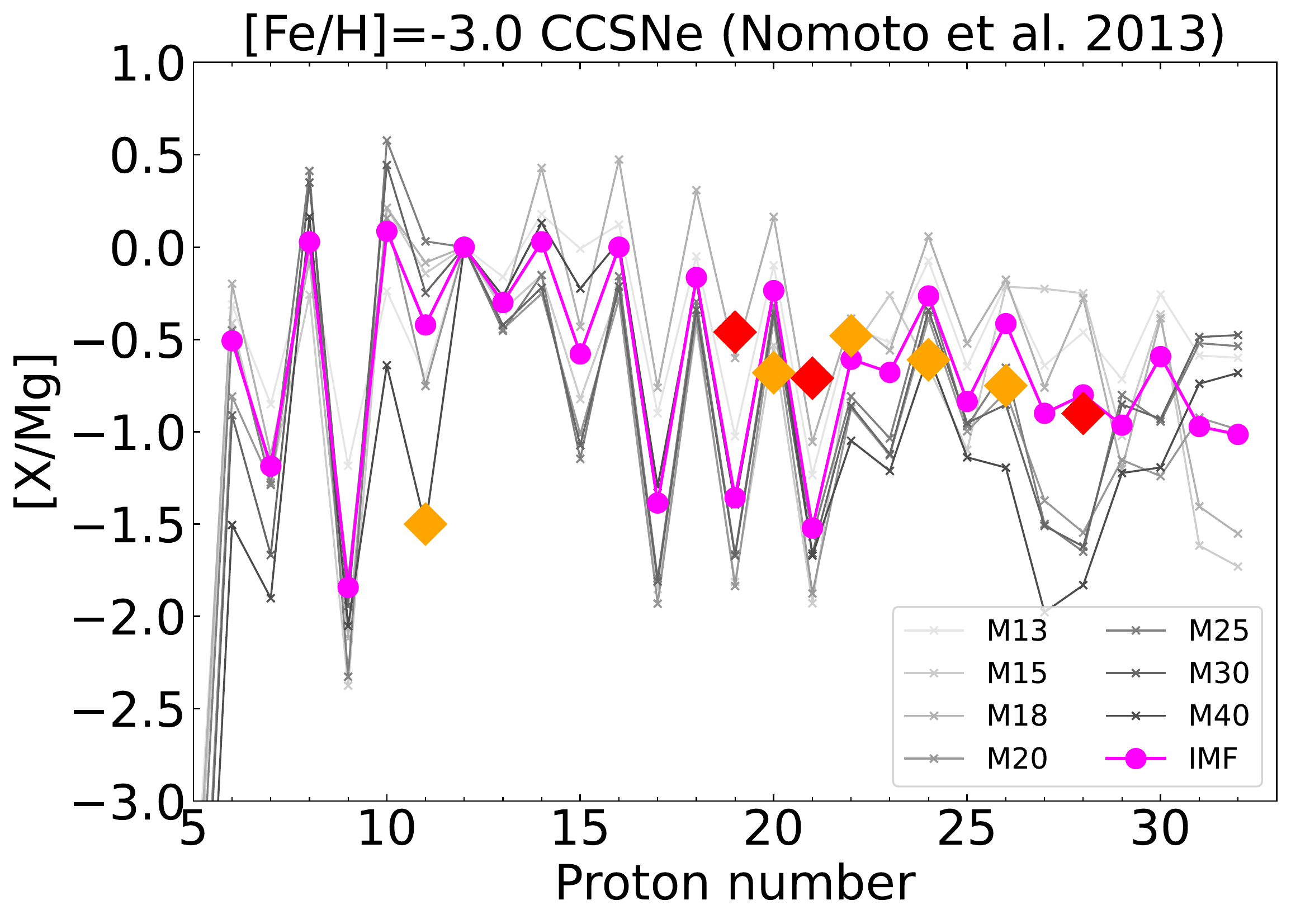}\\
\includegraphics[width=0.50\textwidth]{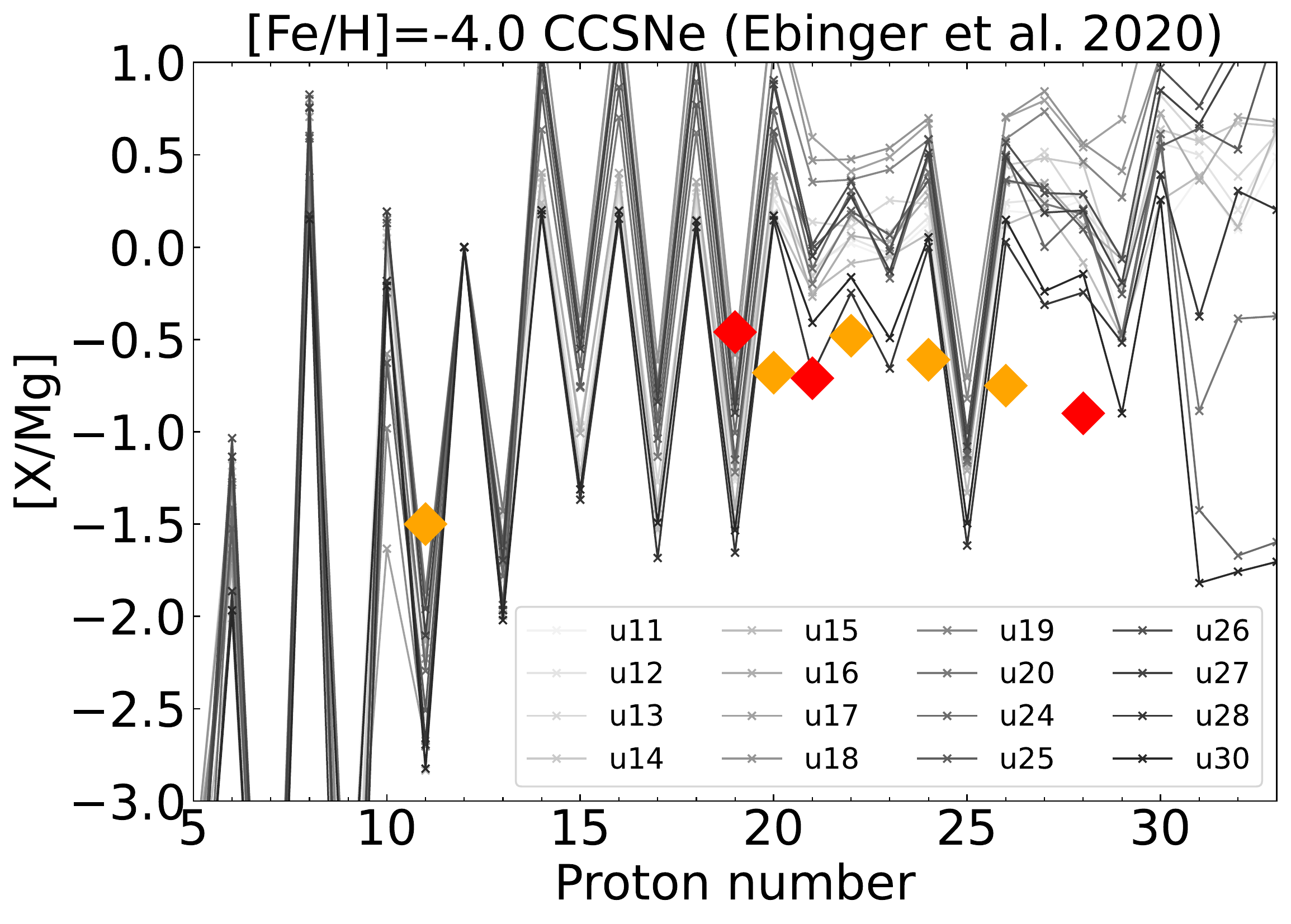}
\caption{Chemistry of Target~1 in the CCSne yields space. Top panel: EMP ($\FeH=-3.0$) CCSNe yields from \citet{Nomoto13}. The ``IMF'' model (magenta line) is the contribution of several SNe~II integrated with a Salpeter initial mass function. The $\chi^2$ is larger by a factor $\sim1.5$ for the IMF fit than the single case SN~II. Bottom panel: UMP ($\FeH=-4.0$) CCSNe from \citet{Ebinger20} in the proton number range as top panel.  The legend indicates the model's name, in which the number is the progenitor's mass in $\msun$ at its ZAMS. The darker the line, the heavier the mass.}
\label{Fig:yields}
\end{figure}

\begin{figure}
\includegraphics[width=0.5\textwidth]{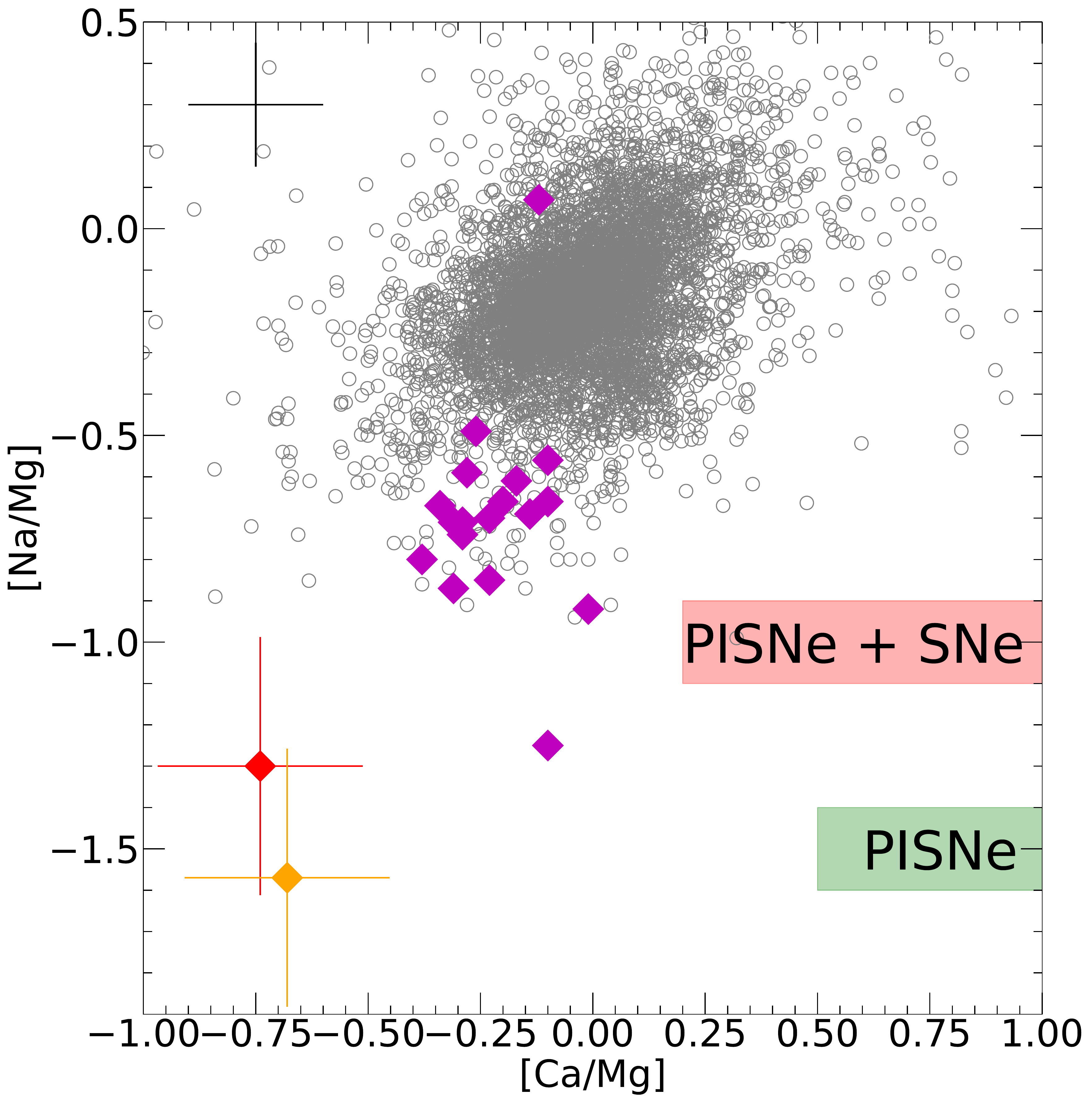}
\caption{PISNe yields space. Target~1 is marked with a red and a orange diamond when LTE and NLTE, respectively. The green band is the region of stars polluted by PISNe alone \citep{Takahashi18}.  The red zone is the locus in which the stars would have been polluted by  PISNe and  SN II as in \citet{Salvadori19}. For the latter case, we show the yields relative to a PISNe to SN II ratio between 0.5 and 0.9 \citep[see Figure~6 from][]{Salvadori19}.
Literature UMi stars (magenta diamonds) from \citet{Shetrone01}, \citet{Sadakane04}, \citet{Cohen10}, \citet{Kirby12}, and \citet{Ural15}. Literature MW halo compilation (grey open circles) from \citet{Aoki13}, \citet{Yong13}, \citet{Kielty21}, and \citet{Buder21}. The black cross at the corner represents the typical  uncertainty on the UMi literature chemical abundances.}
\label{Fig:pisne}
\end{figure}

Various yields of SNe~II are on the market. In Figure~\ref{Fig:yields}, we compare the chemistry of Target~1 against the widely used faint SNe~II yields from \citet[][hereafter NKT13]{Nomoto13} and the more recent yields for ultra metal-poor stars from \citet[][hereafter E20]{Ebinger20}.  Both sets of models are non-rotating, however the yields from E20 reach heavier elements (up to proton number 60, compared to only 32 from NKT13). Another difference is how the energy of the supernovae explosion is parametrized. While NKT13 fixed the energy to the value of $10^{51}$ erg, this is treated as a free parameter by E20, which spans 0.2 to 2.0 $\times10^{51}$ erg, and varies with the progenitor mass. The spatial symmetry of the explosion is also modelled differently. NKT13 employed a mixing and fallback model, which implies the presence of polar jets and fallback material around the equatorial plane, whereas E20 adopted spherical symmetry.  

When comparing the yields from NKT13 with Target~1, the chemistry of this star is sufficiently well described by pollution from a low-mass faint CCSNe ($\lesssim30\msun$). Taking the integrated contribution of several SNe~II over a standard Salpeter IMF provides a worse fit to the Target 1 chemical properties (magenta line in Figure~\ref{Fig:yields}).
This reinforces the hypothesis that Target 1 was born from gas polluted by $\sim$one supernova, as in the ``one shot'' model. A comparison with the yields from E20 for even lower metallicity CCSNe is worse.  Their predictions at all masses are higher for the majority of elements, and the predicted odd-even effect is even stronger.  

Finally, we note that pair-instability supernovae (PISNe) are produced by very metal-poor, very massive stars ($>120\msun$), predicted to be amongst the first stars.  PISNe produce a strong odd-even effect in the yields, with no neutron-capture process elements above the mass cut \citep{Takahashi18}. The odd-even effect leads to high [Ca/Mg]  and low [Na/Mg]. The chemical abundances in Target 1 do not resemble the predictions for PISNe, nor do those for stars in UMi from the literature (see Figure~\ref{Fig:pisne}).

\subsubsection{Presence of rapid- and slow-neutron capture processes}\label{sec:rapidslow}
\begin{figure}
\includegraphics[width=0.475\textwidth]{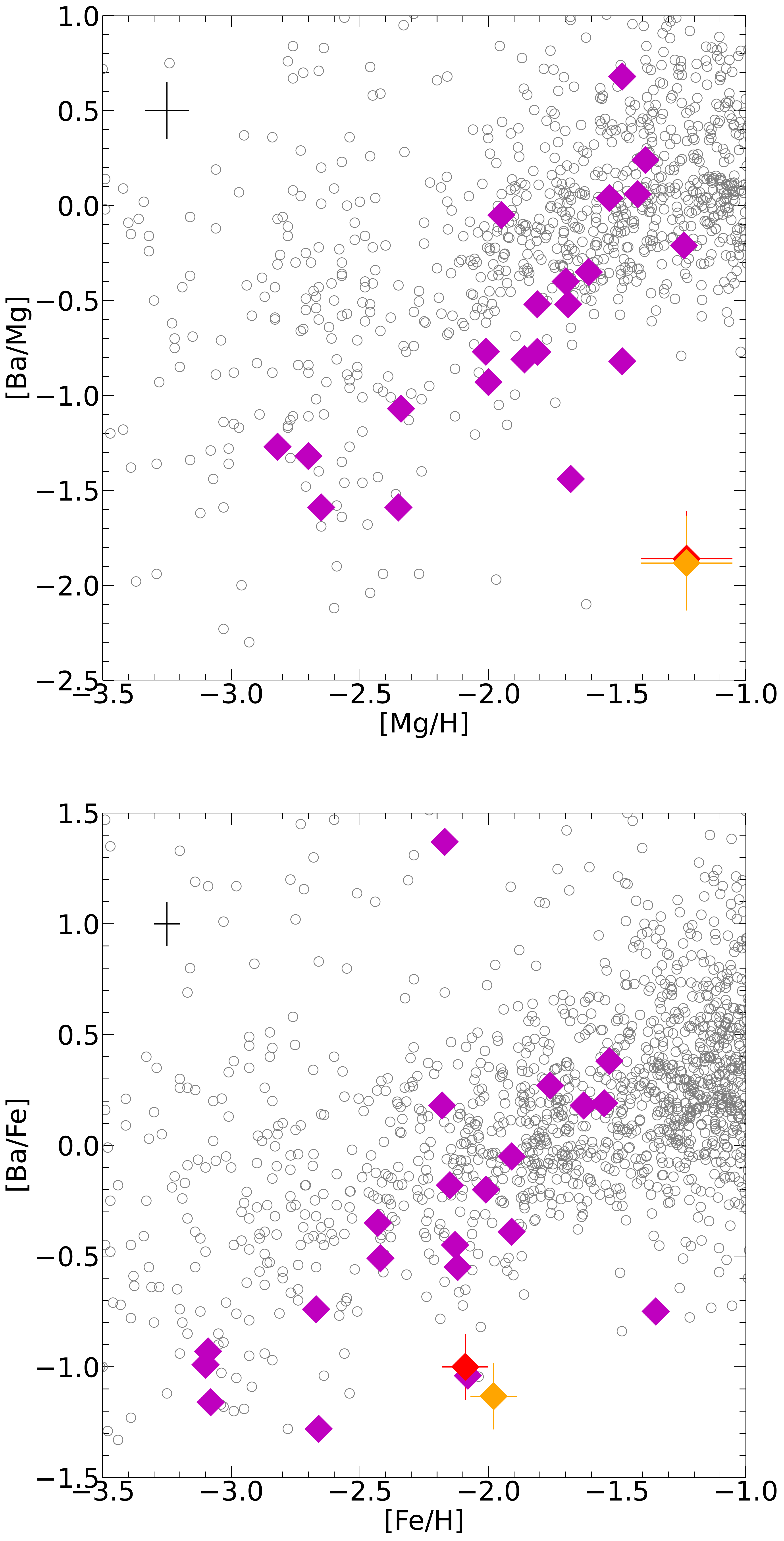}
\caption{ Top panel: [Ba/Mg] vs. [Mg/H] space. Bottom panel: [Ba/Fe] vs. [Fe/H] as in Figure~\ref{Fig:chems}. Target~1 is denoted with a red (LTE) and a orange (NLTE) diamond. Literature UMi stars (magenta diamonds) are from \citet{Shetrone01}, \citet{Sadakane04}, \citet{Cohen10}, \citet{Kirby12}, and \citet{Ural15}. Literature MW halo compilation (grey open circles) from \citet{Aoki13}, \citet{Yong13}, \citet{Kielty21}, and \citet{Buder21}. The black cross at the upper left corner represents the typical  uncertainty on the UMi literature chemical abundances.}
\label{Fig:bamg}
\end{figure}

To examine the contributions from SNe~II in UMi, we further examine the distribution in [Ba/Mg] vs. [Mg/H] in Figure~\ref{Fig:bamg}. At very low-metallicities, if Ba is produced by the r-processes \citep[see the review by][and references therein]{Cowan21}, then a tight and flat distribution will be visible, \ie a Ba-floor, also shown in \citet{Mashonkina22}. This seems to be the case for UMi stars with [Mg/H]$<-2.0$, including Target~1. A spread in [Ba/Mg] that is significantly larger than a 3$\sigma$ error, and subsequent rise from a presumed Ba-floor, is interpreted as Ba contributions from metal-poor asymptotic giant branch stars (AGB), via slow neutron-captures  \citep[s-process, \eg][]{Pignatari08,Cescutti14}. This chemical behaviour is also visible in the bottom panel of Figure~\ref{Fig:bamg}, in which we report the [Ba/Fe] vs. \FeH{} (as in Figure~\ref{Fig:chems}) as a check that our interpretation is not biased by measurements of Mg.  We note that Target~1 clearly separates from the Milky Way population in Figure~\ref{Fig:bamg}, which further validates that this is not a foreground MW star.

Based on an overabundance of [Y/Ba] observed in UMi stars at  very low metallicities, [Fe/H]$<-2.5$, \citet{Ural15} have suggested that there are also contributions from spinstars \citep[\eg][]{Cescutti13} at the earliest epochs. Spinstars are fast rotating massive stars (25--40 $\msun$) that produce s-process elements from neutron rich isotopes in their atmospheres \citep[\eg][]{Cescutti14}. Unfortunately, our GRACES spectra are insufficient (SNR too low for the weak \ion{Y}{II} lines) to determine an abundance for [Y/Ba], including our spectrum of Target~1.

\subsubsection{Search for contributions from SN Ia}\label{sec:chemsevo}

The contribution of SNe~Ia in UMi is still under debate \citep[\eg][and references therein]{Ural15}. The flat distribution in the $\alpha-$ and Fe$-$peak elements shown in Figure~\ref{Fig:chems} are consistent with no contributions from SN Ia, with the exception of the most metal-rich star, COS171  \citep{Cohen10}. While this lone star might draw the eye to the conclusion of a possible $\alpha-$knee (\ie the rapid change in the slope of the  $\alpha-$elements from a plateau to a steep decrease), it is the  [Na, Ni/Fe] (and likely [Ti, Sc/Fe]) ratios that favour the steep decrease, and suggest the presence of contributions from SN Ia. In support, \citet{McWilliam18} analysed COS171 and found that its [Mn, Ni/Fe] ratios do indicate SN~Ia contributions, from sub-Chandrasekhar-mass degenerate white dwarfs, \ie $\sim0.95 \msun$.

\begin{figure}
\includegraphics[width=0.475\textwidth]{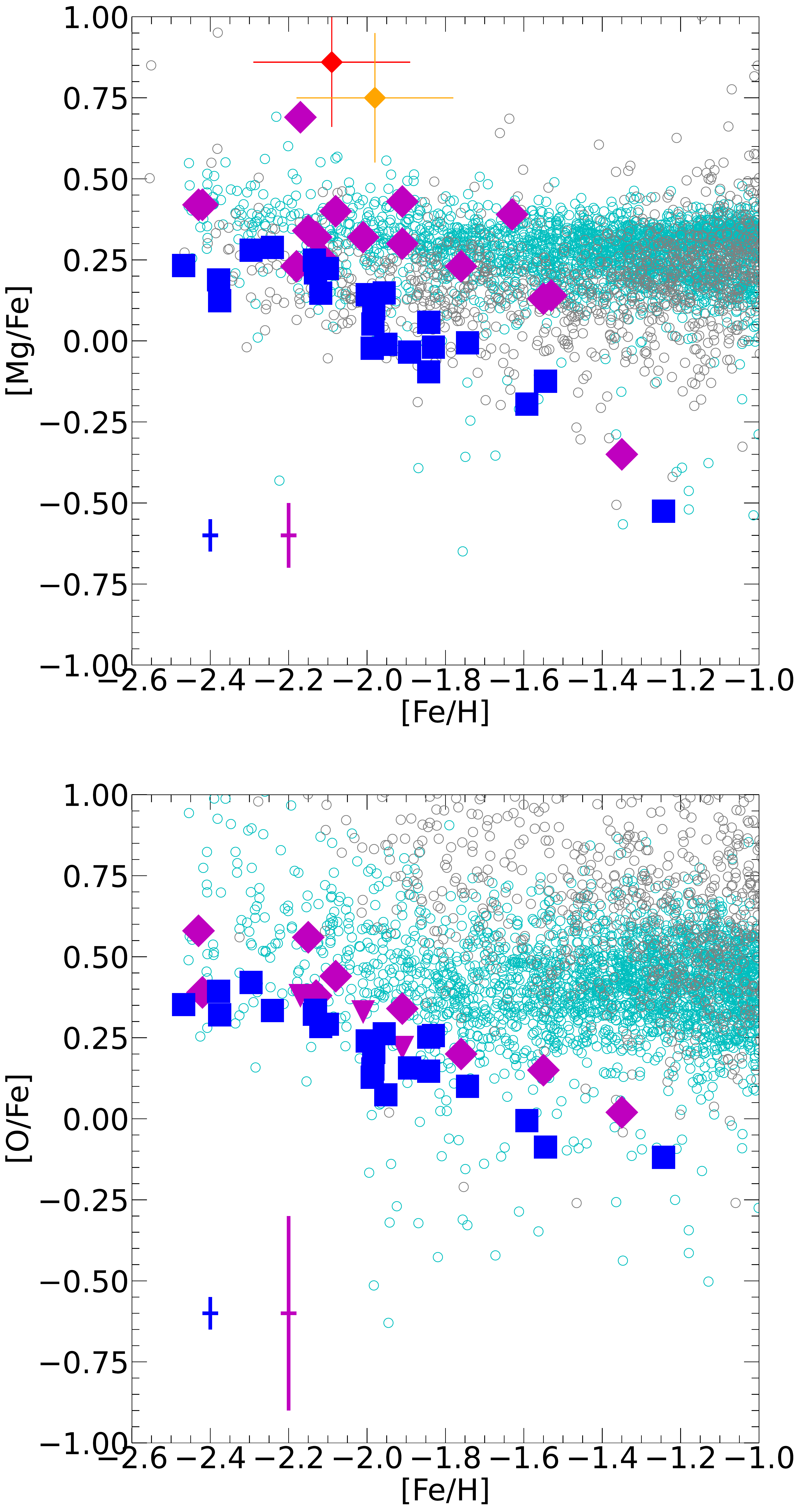}
\caption{UMi chemical abundances from APOGEE DR17 \citep{APOGEEDR17}. Blue squares are stars from APOGEE with high SNR ($>70$) and very likely to be UMi members (Psat$>70$ percent) according to our algorithm. UMi stars from the literature are marked with magenta squares, while magenta triangles denote their upper limits. Target~1 is marked with a red  (LTE) and orange (NLTE) diamond. Cyan open circles are MW stars from APOGEE with high SNR ($>70$) and good Gaia EDR3 parallax measurements ($\varpi/\delta_{\varpi} >15$). Grey open circles are MW stars from GALAH \citep{Buder21} selected as in Figure~\ref{Fig:chems}. Typical uncertainties are denoted with blue and magenta crosses for APOGEE (infrared NLTE) and literature stars (high-resolution optical LTE), respectively. An offset in [Mg/Fe] between the optical LTE and infrared NLTE measurements is under investigation by the APOGEE team (Shetrone et al. 2023, in prep.). }
\label{Fig:knee}
\end{figure}

To further investigate the contributions from SNe~Ia, we explore the data for stars in UMi from APOGEE DR17 \citep{APOGEEDR17}\footnote{\url{https://www.sdss4.org/dr17/irspec/abundances}}, along with data from optical analyses in the literature (see Section~\ref{sec:chemslit}). We compare [Mg/Fe] and [O/Fe] abundances with metallicity in Figure~\ref{Fig:knee} between stars in UMi and stars in the MW halo \citep[from the GALAH survey,][]{Buder21}. Mg and O are amongst the most reliable $\alpha$-element abundance indicators, although O can be challenging in the optical (e.g., very weak [O\ione{}] lines at $\lambda\lambda 6300, 6363$ \AA, or strong [O\ione{}] T lines that suffer from large NLTE effects at $\lambda\lambda 7772,7774,7775$ \AA). With the addition of reliable [O/Fe] from APOGEE, the presence of a plateau up to $\FeH\lesssim-2.1$ followed by a steeper decrease, \ie a knee, is more clearly seen.  APOGEE [Mg/Fe] results also show a steep decrease\footnote{A deeper analysis of the APOGEE spectra in terms of the chemo-dynamical analyses of dwarf galaxies is currently under investigation, Shetrone et al. (2023, in prep.). That study will also explore any offsets between optical and infrared measurements, \eg for [Mg/Fe] as seen in Figure~\ref{Fig:knee}}, indicating the presence of SN Ia contributions.

The metallicity at which the knee occurs ($\FeH_{\rm{knee}}$), is correlated with the time when SNe~Ia begin to contribute to the chemical evolution of a galaxy.  This time is also dependent on the star formation efficiency, which is expected to be lower in dwarf galaxies \citep[\eg][]{Matteucci03,Tolstoy09}. Recently, \citet{Theler20} have suggested that the {\it slope} of the knee-decrease is governed by the balance between the amount of metals ejected by SNe~Ia vs. SNe~II; a smaller slope indicates an extended star formation rather than a sharply quenching galaxy. 

On the theoretical side, \citet{Revaz18} developed cosmological zoom-in simulations that are able to reproduce most of the observable quantities of dwarf galaxies, \eg velocity dispersion profiles, star formation histories, stellar metallicity distributions, and [Mg/Fe] abundance ratios. The \textsc{FIRE} simulations \citep[\eg][]{Hopkins14} have also been used to (a) reproduce the star formation histories of the MW satellites \citep{Escala18}, and (b) reproduce the properties and numbers of ultra-faint dwarf galaxies \citep{Wheeler15}. These models suggest that a higher $\FeH_{\rm{knee}}$ is attained when the star formation is more efficient and the system can retain the metals. Given the value of [Fe/H]$_{\rm knee}\sim-2.1$, then the low star formation efficiency of UMi appears to be similar to measurements in other dwarf galaxies \citep[\eg][]{Reichert20, Tolstoy09, Simon19}, and much less efficient than in the MW, where $\FeH_{\rm{knee}}\sim-0.5$, \citep[\eg][]{Venn04, Haywood2013, Buder21, RecioBlanco22}.

\subsection{An extended "stellar halo" and tidal effects}

Previously, it was shown that  UMi is more elongated ($\epsilon_{\rm UMi}=0.55$) than other classical satellites \citep[$\epsilon<0.45$,][]{Munoz18}. The most distant member had been located near $\sim5.5 r_h$. With our results, Ursa Minor extends out to a projected elliptical distance of $\sim12 r_h$, or $\sim4.5$ kpc (projected) from its centre. This distance is  close to the tidal radius inferred by \citet{Pace20}, $5-6$ kpc. 

\citet{Errani22a} analysed the dynamical properties of many satellites of the MW in terms of their dark matter content and distribution. The authors show that the dynamical properties of UMi are compatible with $\Lambda-$CDM model if tidal stripping effects are taken into account. The finding of a member at $\sim12 r_h$ the multiple apocentric and pericentric passages reinforce the idea that UMi is strongly dominated by tidal stripping. In fact, as shown in the left panel of Figure~\ref{Fig:onsky}, the proper motion of UMi is almost parallel to the semi-major axis of the system. Alternatively, supernovae feedback can play a role in pushing members to the extreme outskirts of their host galaxy. These scenarios have also been proposed to explain the extended structure of Tucana II ultra-faint dwarf galaxy \citep{Chiti21}. The authors discuss a third possible scenario which involves mergers of UFDs. We discuss and rule out the merger hypothesis for UMi in Section~\ref{sec:nomerger}.

\subsubsection{Outside-in star formation vs. late-time merger}\label{sec:nomerger}
\citet{Pace20} measured radial velocities and metallicities of likely UMi members selected from \textit{Gaia} DR2 within 2 half-light radii. They interpreted the spatial distribution of the stars as composed of two populations with different chemo-dynamical properties. A more metal-rich  ($\overline{\FeH} =-2.05\pm 0.03$) kinematically colder ($\sigma_{\rm{RV}} = 4.9\pm 0.8\kms$) and centrally concentrated ($r_h= 221\pm17$ pc) population. And a  metal-poor hotter and more extended ($\overline{\FeH} =-2.29\pm 0.05$, $\sigma_{\rm{RV}} = 11.5\pm 0.9\kms$, $r_h= 374\pm49$ pc) population. \citet{Pace20} discussed that the two metallicity distributions in UMi are much closer than in other dwarf spheroidal galaxies (dSphs) found so far. 

\citet{BenitezLlambay16}, \citet{Genina19}, and \citet{Chung19} have proposed that dwarf-dwarf mergers may be the cause of the multiple populations in dSphs. Therefore, \citet{Pace20} concluded that UMi underwent a late-time merger event between two dwarfs with very similar chemical and physical properties. However, \citet{Genina19} also pointed out that kinematic and spatial information alone are insufficient to disentangle the formation mechanisms of multi-populations. Additional evidence from precise chemical abundances and star formation histories are needed, data that was not included in the study by \citet{Pace20}. 
 
In this paper, we propose an alternative scenario to explain the chemo-dynamical properties of the two populations in Ursa Minor. An outside-in star formation history can also be used to describe the properties of low mass systems, such as dwarf galaxies \citep{Zhang12}. Briefly, the extended metal-poor population ($\FeH\lesssim-2.0$) formed everywhere in the dwarf, such that the relatively younger stars populate the centre of the galaxy at times when SNe~Ia  begin to contribute \citep[\eg][]{Hidalgo13, BenitezLlambay16}. This enhances the metallicity  only in the central region, giving the galaxy a non-linear metallicity gradient. 

In support of our simpler interpretation, the distributions in the chemical elements over a wide range in metallicity suggests a common path amongst the stars in UMi. UMi stars are polluted by low mass CCSNe (\eg their low [Ba/Fe, Mg] and [Na, Ca/Mg]), they show a SNe~Ia knee at $\FeH\sim-2.1$ and a contribution from AGB is also visible in the more metal-rich stars, and they display a low dispersion in [Ca/Mg] from star to star over 2 dex in metallicity.

Furthermore, \citet{Revaz18} used a cosmological zoom-in simulation to show that the kinematics in UMi are consistent with secular heating in the central region of the satellite without invoking late-time mergers. Thus, a more simple scenario of outside-in star formation is consistent with the chemical, structural, and kinematic properties of UMi, and we suggest these do not necessarily require a late-time merger event.

\subsubsection{Tidal perturbations in Ursa Minor}\label{sec:tides}

To examine if the tidal scenario is the main culprit of the extended stellar halo,  candidate members of Ursa Minor from the algorithm of \citet{Jensen23} are used. These have been selected with a total probability (see Section~\ref{sec:selection}) of membership $>30$ percent. To be noted, the algorithm is very efficient in removing foreground contaminants, showing their  probabilities is confined to $\lesssim15$ percent. Additionally, fainter stars than G $=19.5$mag have been removed, since their Gaia proper motion is less reliable. 

\begin{figure}
\includegraphics[width=0.5\textwidth]{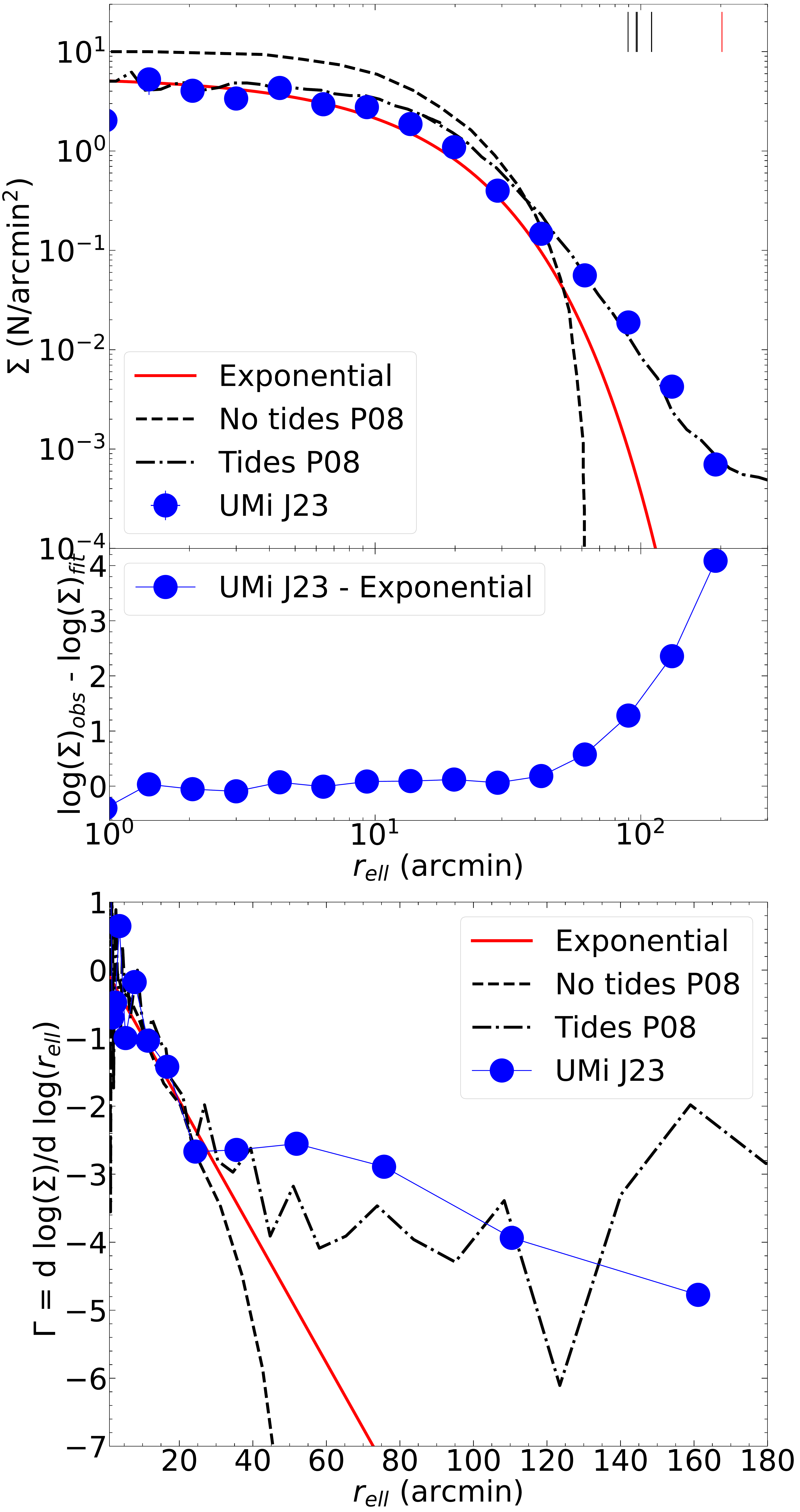}
\caption{Top panel: Surface density distribution $\Sigma$. Central panel: Departure from the exponential fit. Bottom panel: Logarithmic derivative of the surface density, $\Gamma$. UMi candidate members from \citet[][J23]{Jensen23} are marked with blue circles.  Models from \citet[][P08]{Penarrubia08} are denoted with a dashed line (no tide) and with a dash-dotted line (model relative to a first apocentric passage). Red line is the exponential fit to J23 data. Black and red ticks in the top panel mark the position of the 5 Targets (Target~1 in red).}
\label{Fig:surf}
\end{figure}

The surface number density profile $\Sigma$ and its logarithmic derivative $\Gamma$ are derived and compared against numerical simulations. If unperturbed dwarf galaxies are well represented by an exponential profile \citep[\eg][]{McVenn2020a}, the chance of detecting stars as far out as 10 half-light radii from the centre would be negligible. Top and central panels of Figure~\ref{Fig:surf} clearly show that the surface density of the candidate members (blue circles) shows a large excess in the outer regions over an exponential profile (solid red line). Indeed, the surface density at the farthest distance bin is $10^4$ larger than the exponential (central panel).

Is this excess mainly driven by Galactic tidal perturbation? Tides, moving stars to the outer region of a system, would affect the surface number density distribution. Signatures of a tidal perturbation are best recognised in its logarithmic derivative, $\Gamma=\rm{d}\log\Sigma/\rm{d}\log r$,  which is displayed, as a function of radius, in the bottom panel of Figure~\ref{Fig:surf}.  Stars affected by tidal perturbation would gain energy and move outwardly, forming an excess over the exponential profile (\ie less negative slope in this panel).  The sudden departure from the exponential shows as a "kink" in the $\Gamma$ profile, which is  visible in the data at r$\sim30$ arcmin from the centre (blue circles). The $\Gamma$ profile due to tides is expected to approach a power-law of index $-4$, which is a horizontal line of $\Gamma=-4$ \citep[][]{White87,Jaffe87,Penarrubia09}.

These departures from an exponential profile are also visible in the N-body simulations from \citet[][see the top-left panel of their Figure~4]{Penarrubia08}, scaled to the same elliptical distance as the observational "kink" radius. No-tide and tidally perturbed models are shown with the dashed and dot-dashed curves, respectively.  As expected, the tidal model follows the R$^{-4}$ power-law in the outer regions, ending at a "break" radius where the $\Sigma$ profile flattens and $\Gamma$ is expected to show a sudden upturn.  In the simulations, the "break radius" is located at ~200 arcmin, farther still than the position of the outermost candidate member. We emphasize that presence of a "kink" and the power-law profile ($\Gamma=-4$) are not expected in models where the excess stars in the outskirts are a result of gas outflows or mergers \citep[\eg][]{Penarrubia09,BenitezLlambay16}. The presence of a "kink" and a "break" therefore favour a tidal interpretation.

The "break" radius is located where the local crossing time equals the time elapsed since the last pericentric passage. The relation from \citet{Penarrubia09} can then be used to predict the observational "break" radius in UMi\footnote{r$_b~=~C~\cdot~\sigma_{\rm{v}}~\cdot~t_{\rm{peri}}$, where C is a coefficient ($=0.55$),  $\sigma_{\rm{v}}$ is the velocity dispersion of the system ($9.5\kms$, see Table~\ref{tab:umiprop}), and $t_{\rm{peri}}$ Gyr is the time since the last pericentric passage. For the latter, \citet{Battaglia22} provides three values according to their different potentials, $t_{\rm{peri}} = 3.61$ Gyr for the MW$+$LMC, $t_{\rm{peri}} = 2.03$ Gyr for the isolated heavier model, and $t_{\rm{peri}} = 0.93$ Gyr for the  isolated lighter model.}. Adopting the smallest time since the last pericentric passage \citep[$t_{\rm{peri}} = 0.93$ Gyr,][]{Battaglia22}, we find a "break" radius $r_b\sim225$ arcmin ($\sim 5$ kpc or $\sim 13$ r$_h$). This value is close to the tidal radius inferred by \citet[][$5-6$ kpc]{Pace20} and slightly larger than the position of our outermost star. The majority of stars within the "break" radius are bound, which implies that our targets are still bound to the system.

Very recently, \citet{Sestito23scl} discovered a similar excess in the surface density profile of Sculptor and performed the same exercise to test for Galactic tides in the system. They also discuss that this excess is of tidal origin rather than an innate feature of the system. In fact, they tested this scenario for the case of Fornax \citep[][and references therein]{Borukhovetskaya22,Yang22}, a  system largely expected to have been unaffected by tides. The authors find that the surface number density profile of Fornax is well described by an exponential function even at very large distances.  Based on the existence of a "kink" in UMi,  the R$^-4$ outer profile, and the potential existence of a "break" (to be confirmed by extending the profile beyond $\sim200$ arcmin) we conclude that the outer profile of UMi is a result of the effect of Galactic tides.

\section{Conclusions}
A new Bayesian algorithm \citep{Jensen23} was used to find new members in the very extreme outskirts of the dwarf galaxy, Ursa Minor.  Five targets were selected for high-resolution spectroscopy with GRACES at Gemini North. For all five stars, we determine precise radial velocities and metallicities; for the brightest and farthest target in projection (Target~1), the higher SNR of our GRACES spectrum also permitted a detailed chemical abundance analysis.  With the use of data from the literature and APOGEE DR17, we find that:

\begin{enumerate}
    \item The Bayesian algorithm from \citet{Jensen23} is very efficient at finding new members, even at  very large elliptical distances. All five candidates are new members of UMi, according to their radial velocities and metallicities (see Figure~\ref{Fig:rv_met_rh}).
    \item Ursa Minor extends at least out to a projected elliptical distance of $\sim12r_h$, which corresponds to $\sim4.5$ kpc for an adopted distance of $76$ kpc.
    \item The chemical properties of Target~1 (see Figure~\ref{Fig:chems}), the most distant member discovered so far, are compatible with the overall distribution of the known UMi members from high-resolution spectral analysis.
    \item The low [Ca, Na/Mg] and the low [Ba/Fe] of Target~1 suggest that the star formed in an environment polluted by low-mass supernovae type II (M$_{\rm{prog}}\sim30\msun$, see Figures~\ref{Fig:yields}~and~\ref{Fig:pisne}). The star is likely exiled by tidal forces and/or supernovae feedback.
    \item Ursa Minor is also clearly polluted by supernovae type II and AGB stars given the distribution of [Ba/Mg, Fe] as a function of [Mg, Fe/H] (see Figure~\ref{Fig:bamg}).
    \item There is no trace of yields from pair-instability supernovae, either alone or combined with type II (see Figure~\ref{Fig:pisne}).
    \item Looking at all the UMi stars with high-resolution chemical analyses, including those from APOGEE DR17, we conclude there is evidence of pollution by supernovae type Ia. There is a knee at $\FeH_{\rm{knee}}\sim-2.1$ in the [Mg, O, Na, Ni/Fe] distributions (see Figures~\ref{Fig:chems}~and~\ref{Fig:knee}).
    \item The chemo-dynamical properties of UMi can be explained by an outside-in star formation and the following SNe~Ia enrichment. We propose this as a simpler scenario than a late-time merger event between two very similar systems (see Section~\ref{sec:nomerger}). 
    \item The surface density distribution and its logarithmic derivative (see Figure~\ref{Fig:surf}) clearly show that UMi is perturbed by tidal forces starting from a projected distance of $\sim30$ arcmin, the "kink" radius.
    \item The distance of the outermost member is inside the break radius (calculated here as $\gtrsim225$ arcmin), therefore, Target~1  is still bound to UMi.
    \item We find two new UMi members at a distance of $\sim7r_h$ in APOGEE DR17 (Section~\ref{sec:selection} and Figure~\ref{Fig:onsky}). As their metallicities are at the edge of the APOGEE grid ($\sim-2.4$), their true \FeH{} may be  lower and their chemical ratios might be affected.
    
\end{enumerate}

In the very near future, the Gemini High resolution Optical SpecTrograph \citep[GHOST,][]{Pazder16} will be operative at Gemini South. It will cover a wider spectral region than GRACES, especially towards the blue where many spectral lines of heavy elements are found. In synergy with \textit{Gaia} satellite and the powerful Bayesian algorithm for target selections, it should be possible to discover a plethora of new members in the centre and extreme outskirts of this and many other ultra-faint and classical dwarf galaxies to study their star formation histories. This will be a giant leap forward for detailed studies of low mass systems, and both observational and theoretical near field cosmological investigations.

\section*{Acknowledgements}
We acknowledge and respect the l\textschwa\textvbaraccent {k}$^{\rm w}$\textschwa\ng{}\textschwa n peoples on whose traditional territory the University of Victoria stands and the Songhees, Esquimalt and $\ubar{\rm W}$S\'ANE\'C  peoples whose historical relationships with the land continue to this day.

The authors wish to recognize and acknowledge the very significant cultural role and reverence that the summit of Maunakea has always had within the Native Hawaiian community. We are very fortunate to have had the opportunity to conduct observations from this mountain.

The authors thank the anonymous referee for their precious feedback that improved the quality of the manuscript.

We want to thank the supporter astronomers, Joel Roediger and Hyewon Suh, for their help during Phase II and the observational runs.

FS thanks the Dr. Margaret "Marmie" Perkins Hess postdoctoral fellowship for funding his work at the University of Victoria. KAV, LDA, and JG thank the National Sciences and Engineering Research Council of Canada for funding through the Discovery Grants and CREATE programs. DZ thanks the Mitacs Globalink program for summer funding.  The authors thanks the International Space Science Institute (ISSI) in Bern, Switzerland, for funding the  "The Early Milky Way" Team led by Else Starkenburg.

Based on observations obtained through the Gemini Remote Access to CFHT ESPaDOnS Spectrograph (GRACES), as part of the Gemini Program GN-2022A-Q-128. ESPaDOnS is located at the Canada-France-Hawaii Telescope (CFHT), which is operated by the National Research Council of Canada, the Institut National des Sciences de l’Univers of the Centre National de la Recherche Scientifique of France, and the University of Hawai’i. ESPaDOnS is a collaborative project funded by France (CNRS, MENESR, OMP, LATT), Canada (NSERC), CFHT and ESA. ESPaDOnS was remotely controlled from the international Gemini Observatory, a program of NSF’s NOIRLab, which is managed by the Association of Universities for Research in Astronomy (AURA) under a cooperative agreement with the National Science Foundation on behalf of the Gemini partnership: the National Science Foundation (United States), the National Research Council (Canada), Agencia Nacional de Investigaci\'{o}n y Desarrollo (Chile), Ministerio de Ciencia, Tecnolog\'{i}a e Innovaci\'{o}n (Argentina), Minist\'{e}rio da Ci\^{e}ncia, Tecnologia, Inova\c{c}\~{o}es e Comunica\c{c}\~{o}es (Brazil), and Korea Astronomy and Space Science Institute (Republic of Korea).

This work has made use of data from the European Space Agency (ESA) mission
{\it Gaia} (\url{https://www.cosmos.esa.int/gaia}), processed by the {\it Gaia}
Data Processing and Analysis Consortium (DPAC,
\url{https://www.cosmos.esa.int/web/gaia/dpac/consortium}). Funding for the DPAC
has been provided by national institutions, in particular the institutions
participating in the {\it Gaia} Multilateral Agreement.

Funding for the Sloan Digital Sky Survey IV has been provided by the Alfred P. Sloan Foundation, the U.S. Department of Energy Office of Science, and the Participating Institutions. 

SDSS-IV acknowledges support and resources from the Center for High Performance Computing  at the University of Utah. The SDSS website is www.sdss4.org.

SDSS-IV is managed by the Astrophysical Research Consortium for the Participating Institutions of the SDSS Collaboration including the Brazilian Participation Group, the Carnegie Institution for Science, Carnegie Mellon University, Center for Astrophysics | Harvard \& Smithsonian, the Chilean Participation Group, the French Participation Group, Instituto de Astrof\'isica de Canarias, The Johns Hopkins University, Kavli Institute for the Physics and Mathematics of the Universe (IPMU) / University of Tokyo, the Korean Participation Group, Lawrence Berkeley National Laboratory, Leibniz Institut f\"ur Astrophysik Potsdam (AIP),  Max-Planck-Institut f\"ur Astronomie (MPIA Heidelberg), Max-Planck-Institut f\"ur Astrophysik (MPA Garching), Max-Planck-Institut f\"ur Extraterrestrische Physik (MPE), National Astronomical Observatories of China, New Mexico State University, New York University, University of Notre Dame, Observat\'ario Nacional / MCTI, The Ohio State University, Pennsylvania State University, Shanghai Astronomical Observatory, United Kingdom Participation Group, Universidad Nacional Aut\'onoma de M\'exico, University of Arizona, University of Colorado Boulder, University of Oxford, University of Portsmouth, University of Utah, University of Virginia, University of Washington, University of Wisconsin, Vanderbilt University, and Yale University.

This research has made use of the SIMBAD database, operated at CDS, Strasbourg, France \citep{Wenger00}. This work made extensive use of \textsc{TOPCAT} \citep{Taylor05}.

\section*{Data Availability}
GRACES spectra will be available at the Gemini Archive web page \url{https://archive.gemini.edu/searchform} after the proprietary time. The data underlying this article are available in the article and upon reasonable request to the corresponding author.

\bibliographystyle{mnras}
\bibliography{umi_graces}

\appendix

\section{Orbital parameters for UMi}\label{sec:orbpar}
In this section, we want to test the gravitational potential so far used for kinematical studies in the  disk and the halo of the Milky Way \citep[\eg][]{Sestito19,Sestito20,Lucchesi22}. We make use of \textsc{Galpy}\footnote{\url{http://github.com/jobovy/galpy}}  \citep{Bovy15} to infer the pericentric, apocentric, and galactocentric distances of Ursa Minor. The choice on the isolated gravitational potential and on all the other assumptions (\eg distance and motion of the Sun etc.), the orbital integration time, and the derivation of the uncertainties mirror the method fully described in \citet{Sestito19}. The code is run on the sample of stars from \citet{Spencer18}, \citet{Pace20}, and our five new targets. The system's orbital parameters are obtained from the median of the sample. The uncertainties on the system parameters are derived dividing the dispersion by the square root of the number or stars in the sample. The inferred quantities are compared with the values from the literature \citep{LiH21,Battaglia22,Pace22}, in which a variety of MW gravitational potentials were adopted. In particular, \citet{LiH21} make use of four isolated MW gravitational potential, one NFW dark matter halo (\textsc{PNFW}) and three with Einasto profiles (\textsc{PEHM}, \textsc{PEIM}, and \textsc{PELM}). \citet{Battaglia22} adopted two isolated MW profiles (\textsc{LMW} and \textsc{HMW}) and one perturbed by the the passage of the Large Magellanic Cloud (\textsc{PMW}). \citet{Pace22} used two gravitational potentials, one in which the  MW is isolated (\textsc{MW}),  and the other perturbed by the LMC (\textsc{MW+LMC}). Both \citet{Battaglia22} and \citet{Pace22} make use of NFW dark matter profiles.

\begin{figure}
\includegraphics[width=0.48\textwidth]{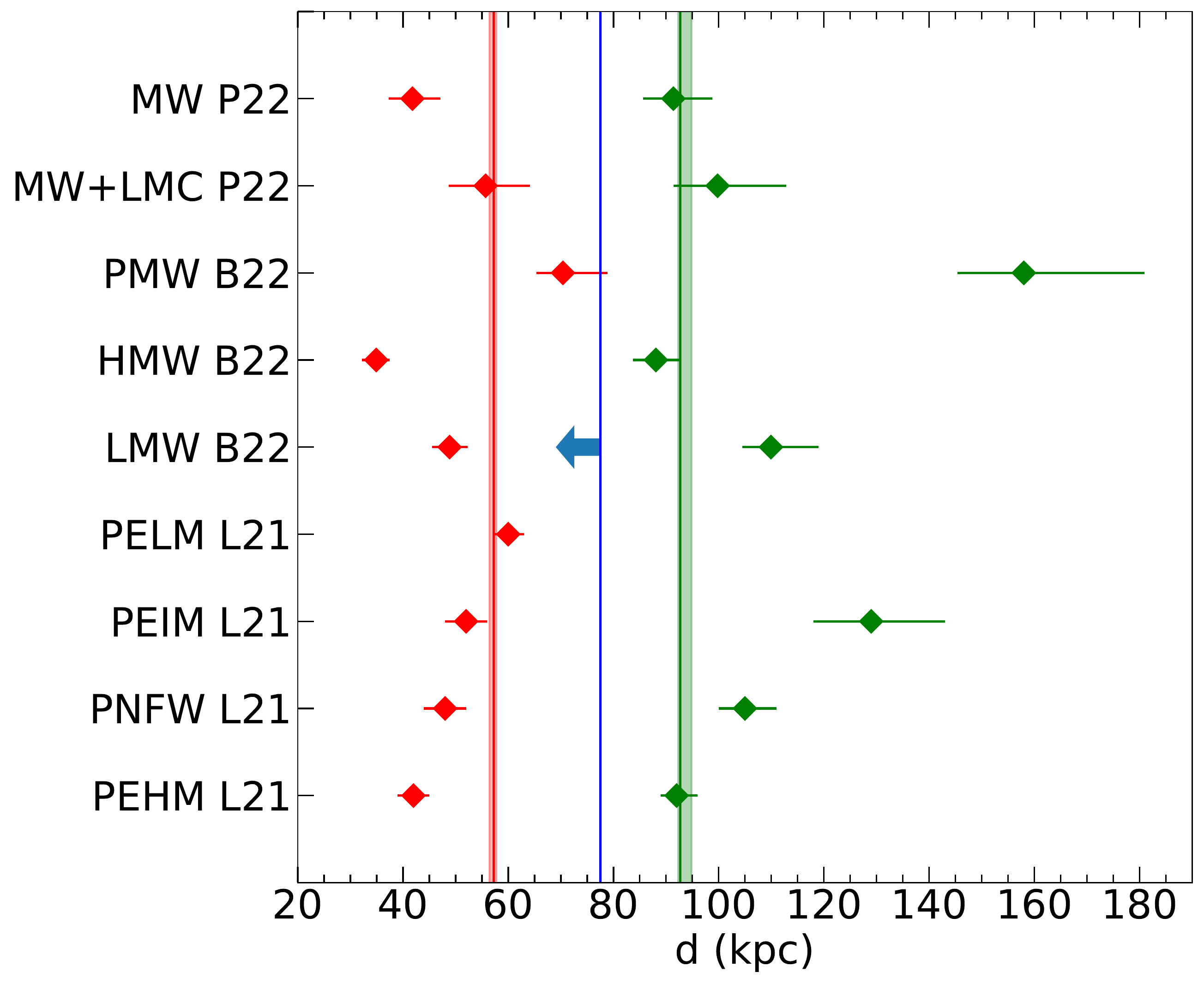}
\caption{Orbital parameters for Ursa minor. The red, green, and blue vertical bands are the pericentric ($R_{\rm{peri}} = 57.23_{-0.83}^{+0.48}$ kpc), apocentric ($R_{\rm{apo}} = 92.67_{-0.41}^{+2.17}$ kpc), and Galactocentric distances ($R_{\rm{GC}} = 77.55_{-0.03}^{+0.02}$ kpc) inferred in this work. To infer the orbital parameters, we use the \citet{Spencer18,Pace20} compilation. Vertical lines are their median values, while shaded area are the interval between the 0.16 and 0.84 quantiles. The blue horizontal arrow departing from the vertical line of the Galactocentric distance represents the direction of the Galactocentric radial velocity. Pericentric  and apocentric distances from the literature are represented by red and red green, respectively. 
 Tick labels in the y axis indicate the studies from which the potentials have been taken, including:  L21  \citep{LiH21}, B22 \citep{Battaglia22}, and P22   \citep{Pace22}.}
\label{Fig:apoperi}
\end{figure}

In Figure~\ref{Fig:apoperi}, we show a range of orbital parameters for UMi -- this is an update to the original results shown by \citet[][their Figure 7]{MartinezGarcia23} for a range of gravitational potentials.
We add our inferred orbits with uncertainties (shaded areas) rather than the median of the literature values.
The Galactocentric position of UMi is closer to its apocentre, yet the blue arrow indicates the system is moving towards its pericentre. 
The inferred orbital parameters are in broad agreement with the results from the variety of gravitational potentials adopted in the literature so far. In particular, the apocentre ($R_{\rm{apo}} = 92.67_{-0.41}^{+2.17}$ kpc) is similar to the ones inferred assuming a more massive dark matter halo, such as the \textsc{PEHM} from \citet{LiH21}, \textsc{HMW} from \citet{Battaglia22}, or \textsc{MW+LMC} and \textsc{MW} from \citet{Pace22}. While the pericentric distance  ($R_{\rm{peri}} = 57.23_{-0.83}^{+0.48}$ kpc) is very different from the one inferred with the \textsc{PMW} from \citet{Battaglia22}, \textsc{PEIM} and \textsc{PELM} from \citet{LiH21}. The pericentre variation is narrower among different potentials, although we can observe our inference is much less in agreement with  \textsc{HMW} and \textsc{PMW} from \citet{Battaglia22}.

\label{lastpage}

\bsp	

\end{document}